\documentclass[fleqn,usenatbib]{mnras}

\usepackage{newtxtext,newtxmath}
\usepackage[T1]{fontenc}
\DeclareRobustCommand{\VAN}[3]{#2}
\let\VANthebibliography\thebibliography
\def\thebibliography{\DeclareRobustCommand{\VAN}[3]{##3}\VANthebibliography}

\usepackage{graphicx}	% Including figure files
\usepackage{amsmath}	% Advanced maths commands
\usepackage{newtxtext,newtxmath}
\newcommand{\Hii}{H~{\sc ii}}
\newcommand{\Hi}{H~{\sc i}}
\newcommand{\cmc}{cm$^{-3}$}

\newcommand{\kms}{km\,s$^{-1}$}
\newcommand{\specialcell}[2][c]{%
	\begin{tabular}[#1]{@{}c@{}}#2\end{tabular}}
\title[A multi-line study of the IRDC G351.78-0.54]{A multi-line study of the filamentary infrared dark cloud G351.78-0.54}

\author[O. L. Ryabukhina and I. I. Zinchenko]{O. L. Ryabukhina$^{1,2}$\thanks{E-mail: ryabukhina@ipfran.ru}
and I. I. Zinchenko$^{1}$
\\
$^{1}$Institute of Applied Physics of the Russian Academy of Sciences, Nizhny Novgorod, Russia \\
$^{2}$Institute of Astronomy of the Russian Academy of Sciences, Moscow, Russia
}

\date{Accepted XXX. Received YYY; in original form ZZZ}

\pubyear{2021}

\begin{document}
\label{firstpage}
\pagerange{\pageref{firstpage}--\pageref{lastpage}}
\maketitle

\begin{abstract}
 We present results of a multi-line study of the filamentary infrared dark cloud  G351.78-0.54 in the 1.3 and 0.8 mm wavelength bands. The lines of the three isotopologues of carbon monoxide CO, N$_2$H$^+$, CH$_3$CCH and HNCO were observed. The aim was to study the general structure of the filamentary cloud, its fragmentation and physical parameters with the emphasis on properties of dense clumps in this cloud. Several dense clumps are identified from the N$_2$H$^+$ (3--2) data, their masses and virial parameters are determined using the C$^{18}$O (2--1) line. Temperatures of some clumps are estimated from the CH$_3$CCH and HNCO data. Almost all clumps appear to be gravitationally unstable. The density estimates obtained from the C$^{18}$O (3--2)/(2--1) and N$_2$H$^+$ (3--2)/(1--0) intensity ratios are in the range $n \sim (0.3-3)\times 10^5$~\cmc. The HNCO emission is detected exclusively toward the first clump which contains the luminous IR source IRAS 17233-3606, and indicates an even higher density. It is observed in the outflow, too. The velocity shift of the higher excitation HNCO lines may indicate a movement of the hot core relative the surrounding medium. In some clumps there is a velocity shift $\sim 1$~\kms\ between N$_2$H$^+$ (3--2) and CO isotopologues. The large widths of the N$_2$H$^+$ (3--2) line in the clumps indicate an increase of the velocity dispersion in their dense interiors, which may be related to the star formation process. The N$_2$H$^+$ abundance drops toward the luminous IR source.
\end{abstract}


%%%%%%%%%%%%%%%%%%%%%%%%%%%%%%%%%%%%%%%%%%%%%%%%%%

%%%%%%%%%%%%%%%%% BODY OF PAPER %%%%%%%%%%%%%%%%%%

\section{Introduction}

Filamentary structures in the interstellar medium are known for a long time already. They have been observed in the far infrared range surveys \citep[e.g.][]{Low84}, in the \Hi\ emission \citep[e.g.][]{McClure06} and in molecular lines \citep[e.g.][]{Bally87,Myers09}. They attracted an enhanced attention since the \textit{Herschel} Space Observatory \citep{Pilbratt10} demonstrated the ubiquitous presence of filaments in nearby clouds \citep{Andre10,Andre2014}. Pre-stellar cores are observed predominantly along the filaments. As a result, a new paradigm of star formation was suggested, in which filaments play a leading role in this process \citep{Andre2014}.

While \textit{Herschel} observations were focused on nearby clouds, extensive galactic-wide surveys of filamentary structures have been performed to date \citep[e.g.][]{Li16,Schisano20}. Statistical distributions of basic physical parameters of interstellar filaments have been derived. Nevertheless, detailed studies of individual objects are required for better understanding their physical properties and processes of star formation. In this respect, infrared dark clouds (IRDC) attract special attention. At least some of them may represent birth places of massive stars \citep{Rathborne06,Kauffmann2010}. The process of high mass star formation is still puzzling in many respects and is actively investigated \citep[e.g.][]{Tan14,Motte18}. Many IRDCs are filamentary \citep{Schisano20}. 

Reliable estimates of the physical conditions in interstellar clouds require multi-transitional data. Here 
we present the results of such multi-transitional study of the filamentary infrared dark cloud G351.78--0.54. It is worth noting that several designations are used for this object in the literature: G351.77-0.54 \citep{Beuther2017}, G351.77–0.51 \citep{2011A&A...533A..85L}, G351.776-0.527 \citep{2019A&A...621A.130L}. In the middle part of this cloud the luminous IR source IRAS 17233--3606 ($\alpha$(J2000) = 17$^{\rm h}$26$^{\rm m}$42\rlap.$^{\rm s}$8, $\delta$(J2000) =  --36$^\circ$09$'$17$''$) is located. This source was a target for many studies \citep{Beuther2017, Beuther2019, Antyufeyev2016, Klaassen2015, Leurini2008, Leurini2011, Leurini2014}. The distance to this object is estimated in the range of 0.7 -- 1 kpc \citep{2011A&A...533A..85L, 2015A&A...579A..91W}. We adopt here the distance of 1~kpc. Investigations of the filament itself are more limited. The length of the filament is 4.6 parsecs, the width is 0.2 pc, the mass exceeds 1300 M$_\odot$  \citep{2019A&A...621A.130L}.  In \cite{2011A&A...533A..85L}   twelve clumps were identified and described with the help of the CLUMPFIND method on the basis the 870 $\mu$m continuum emission and lines of the CO isotopologues. The authors conclude that star formation is continuous at different stages of evolution. The broad profiles of molecular lines can be a consequence of shock compression due to the expanding \Hii\ regions. However, it was not possible to determine the driving source  \citep{2011A&A...533A..85L}.  

\begin{figure*}
	\center{\includegraphics[width=1\linewidth]{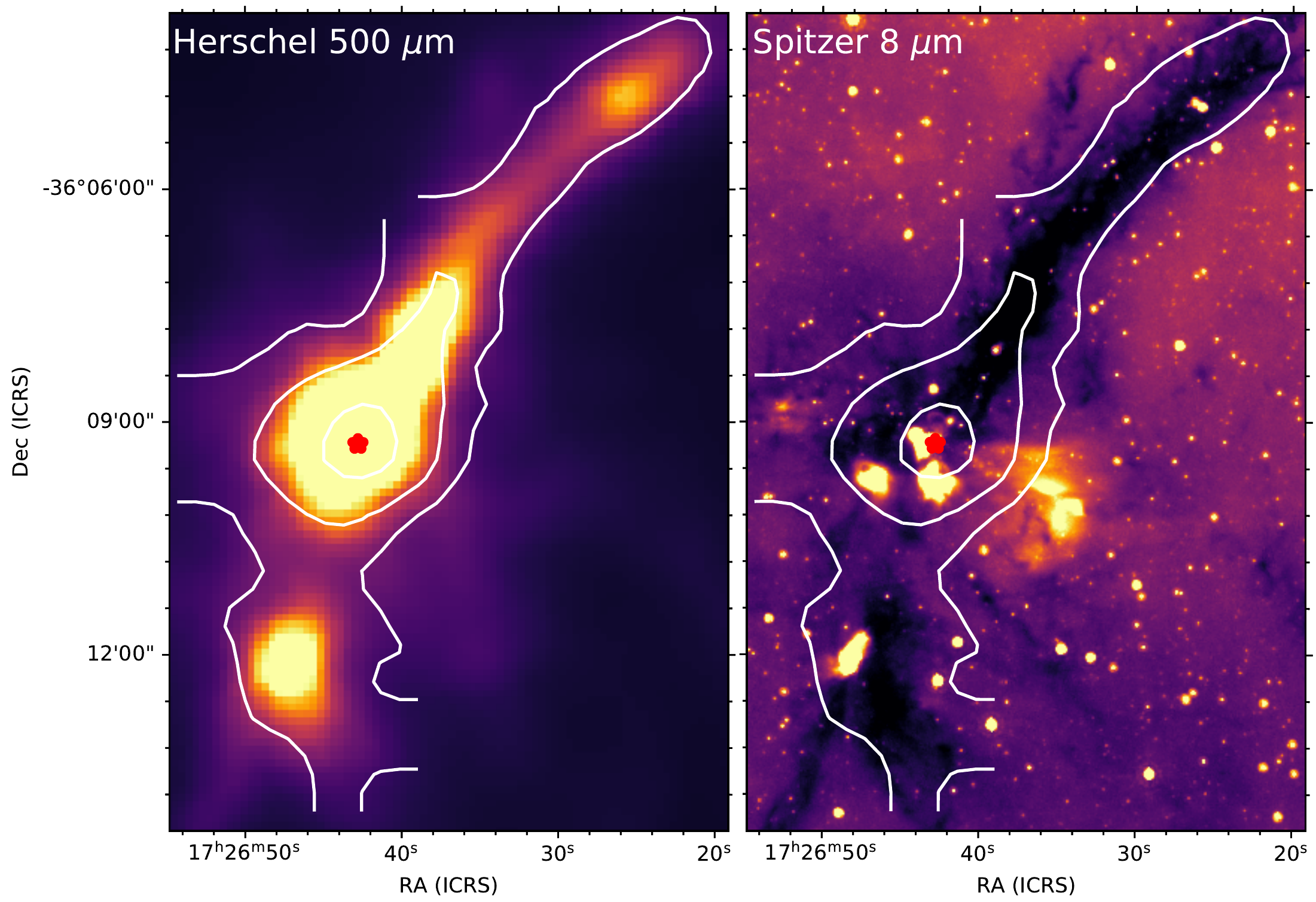}}
	\caption{The map of the region G351.78--0.54 in the infrared range at $\lambda$ = 500~$\mu$m (Herschel) and 8~$\mu$m (Spitzer). The white contours show the C$^{18}$O (2--1) integrated line intensity from our data. Contour levels are 5, 10, 30 K \kms. The position of the IR source IRAS 17233--3606 is marked with an asterisk (red).} 
	\label{g351_ik}
\end{figure*}

Fig. \ref{g351_ik} shows this region in the infrared range at $\lambda$ = 500~$\mu$m (Herschel) and 8~$\mu$m (Spitzer). The filamentary structure is clearly visible in emission at 500~$\mu$m, and as a dark lane in the 8~$\mu$m map. There is a network of fibers associated with the main structure \citep{2019A&A...621A.130L}. 

The aim of the current work is to investigate further the fragmentation and physical properties of this cloud with the emphasis on properties of dense clumps. For this purpose we observed this area in several molecular lines. Here, we present the observations and analysis of the data, including dense clump identification and determination of their properties.

\section{Observations and data analysis} \label{sec:obs}
The observations were carried out with the APEX radio telescope \citep{2006A&A...454L..13G} in 2010--2017 in the wavelength ranges of 1.3 mm and 0.8 mm (projects O-085.F-9323, O-086.F-9316, O-097.F-9303, O-098.F-9306) using SHeFI receivers \citep{2008A&A...490.1157V, 2006SPIE.6275E..0GB}. The beam width at the half power level (HPBW) ranged from 27$^{\prime\prime}$ to 17$^{\prime\prime}$. The data are converted to the main beam temperature scale using the main beam efficiencies ($\eta_{mb}$) presented on the APEX web site. We adopt $\eta_{mb}=0.75$ at 1.3~mm and $\eta_{mb}=0.7$ at 0.8~mm as derived from Jupiter observations since the size of the observed compact structures is comparable to Jupiter. At 0.8~mm there is a rather large scatter in the estimates of this parameter. The adopted value is in a reasonable agreement with the most of these estimates and with the Ruze formula.

The list of the observed lines, which are analyzed here, is presented in Table~\ref{table:lines}. It includes CO (2--1), $^{13}$CO (2--1), C$^{18}$O (2--1), $^{13}$CO (3--2), 	C$^{18}$O (3--2),  N$_2$H$^+$ (3--2), CH$_3$CCH ($13_K-12_K$) and several HNCO transitions. The line parameters are taken mainly from the Cologne Database for Molecular Spectroscopy (CDMS) \citep{Mueller01,Mueller05,Endres16} and from the Jet Propulsion Laboratory (JPL) catalog\footnote{ https://spec.jpl.nasa.gov} \citep{Pickett98}. The spectral resolution (channel width) was 244~kHz in the $^{13}$CO (2--1), C$^{18}$O (2--1), N$_2$H$^+$ (3--2) and HNCO observations. In the CO (2--1), $^{13}$CO (3--2), C$^{18}$O (3--2) and CH$_3$CCH observations it was 76~kHz. The corresponding resolution in velocity is 0.10~\kms\ in CO (2--1) and CH$_3$CCH, 0.33~\kms\ in $^{13}$CO (2--1), C$^{18}$O (2--1) and HNCO (10--9), 0.07~\kms\ in $^{13}$CO (3--2), C$^{18}$O (3--2) and HNCO (15--14), 0.26~\kms\ in N$_2$H$^+$ (3--2).

The median value of the RMS noise at this resolution at the $T_\mathrm{mb}$ scale is 0.13~K in the CO (2--1) spectra, $\sim 0.1$~K in the $^{13}$CO (2--1) and C$^{18}$O (2--1) spectra, $\sim 0.6$~K in the $^{13}$CO (3--2) and C$^{18}$O (3--2) spectra, $\sim 0.1$~K in the N$_2$H$^+$ (3--2) spectra, 0.02--0.11~K in the CH$_3$CCH spectra, $\sim 0.07$~K in the HNCO (10--9) spectra and $\sim 0.09$~K in the HNCO (15--14) spectra.

The observations were performed in the position-switching mode. The reference positions were different for different observations. In some cases a significant emission was apparently present at the reference position resulting in negative features in the measured spectra. In particular this is a case for the $^{13}$CO (2--1) and C$^{18}$O (2--1) observations. At least a part of these data contains a narrow such feature at --3.6~\kms. Its width is $\sim 0.6$~\kms\ in C$^{18}$O (2--1) and $\sim 1$~\kms\ in $^{13}$CO (2--1). The integrated intensity of this feature in the average spectra is $\sim 0.3$~K\,\kms\ in C$^{18}$O (2--1) and $\sim 2$~K\,\kms\ in $^{13}$CO (2--1). For C$^{18}$O (2--1) it constitutes about 4\% of the integrated intensity of the average spectrum and therefore has a negligible effect on the estimates of the C$^{18}$O column density. However this feature should be taken into account when analyzing the line profiles.
In the CO (2--1) spectra there is a negative feature at about --7~\kms\ with the amplitude of about 2.5~K and also several features around --20~\kms. All these features are far from the systemic velocity of the investigated object.
%There is some emission at the reference position in the CO (2--1), $^{13}$CO (2--1), C$^{18}$O (2--1) and  N$_2$H$^+$ (3--2) lines which gives an additional dip in the spectra. For lines $^{13}$CO and C$^{18}$O this dip is at the velocity V$_{\rm LSR}$ is --3.6 \kms, its intensity is 7.7 and 1.1 K respectively, FWHM is 1.2 and 0.9 \kms. In CO line the dip V$_{\rm LSR}$ is --5.2 \kms, intensity is 6.8 K, FWHM is 4.8 \kms, in  N$_2$H$^+$ V$_{\rm LSR}$ is --2.6 \kms, intensity is 0.5 K, FWHM is 5.6 \kms. These dips do not greatly affect the estimates of the integral intensity, but can affect the line profiles.

The line data were processed using the CLASS program from the GILDAS\footnote{http://www.iram.fr/IRAMFR/GILDAS} \citep{Maret2011} package, and for further analysis of the obtained images, the MIRIAD \citep{Sault1995} and Astropy \citep{2018AJ....156..123A} packages were used, Starlink software package \citep{Currie2014} was used to identify molecular clumps. Estimates of the rotational temperature for the CH$_3$CCH transitions were obtained with the CASSIS software\footnote{http://cassis.irap.omp.eu}. The package PySpecKit \citep{2011ascl.soft09001G} was used to simulate the hyperfine structure of the N$_2$H$^+$ (3--2) line. For the excitation analysis of $^{13}$CO, C$^{18}$O and N$_2$H$^+$, the Radex program was used \citep{2007A&A...468..627V}. 

\begin{table}
	\caption{List of observed molecules}
	\label{table:lines}
	\bigskip
	\begin{tabular}{cccc}
		\hline
		Molecule & Transition  &  Frequency & E$_l$  \\
		&& (GHz) &(K)\\
		\hline
		CO & 2--1 & 230.538 & 5.53 \\
		$^{13}$CO &2--1 & 220.399 & 5.29 \\
		 &3--2 &  330.588 & 15.87  \\
		C$^{18}$O &2--1  &219.560  & 5.27\\
		& 3--2 & 329.331 &15.80\\
		N$_2$H$^+$ & 3--2 & 279.512 & 13.41\\
		CH$_3$CCH & 13$_0$--12$_0$ & 222.166 & 63.98  \\
		& 13$_1$--12$_1$ & 222.163 & 71.18 \\
		 & 13$_2$--12$_2$ & 222.150 & 92.78 \\
		 & 13$_3$--12$_3$ & 222.129 & 128.77 \\
		& 13$_4$--12$_4$ & 222.099 & 179.17 \\
		 & 13$_5$--12$_5$ & 222.061 & 243.97 \\
		& 13$_6$--12$_6$ & 222.014 & 323.17 \\
		HNCO  & 10$_{0,10}$--9$_{0,9}$ & 219.798 & 47.47  \\
		& 10$_{1,10}$--9$_{1,9}$ & 218.981 & 90.57 \\
		& 10$_{1,9}$--9$_{1,8}$ & 220.585 & 90.92 \\
		& 10$_{2,9}$--9$_{2,8}$ & 219.734 & 217.74  \\
		& 10$_{2,8}$--9$_{2,7}$ & 219.737 & 217.74  \\
		& 10$_{3,8}$--9$_{3,7}$ & 219.657 & 422.42  \\
		& 10$_{3,7}$--9$_{3,6}$ & 219.657 & 422.42  \\
	    & 15$_{0,15}$--14$_{0,14}$ &  329.664 & 110.76 \\
%		& 15$_1$--14$_1$ &  330.849 & 154.43 \\
		& 15$_{2,14}$--14$_{2,13}$ &  329.573 & 281.01 \\
		& 15$_{2,13}$--14$_{2,12}$ &  329.585 & 281.01 \\
		& 15$_{3,13}$--14$_{3,12}$ &  329.460 & 485.67 \\
		& 15$_{3,12}$--14$_{3,11}$ &  329.460 & 485.67 \\
		& 15$_{4,12}$--14$_{4,11}$ &  329.295 & 761.39 \\
		& 15$_{4,11}$--14$_{4,10}$ &  329.295 & 761.39 \\
		\hline
	\end{tabular}
\end{table}

\section{Results}

\subsection{Gas distribution and kinematics}\label{sect_gasdist}

\begin{figure*}
	\includegraphics[width=1\linewidth]{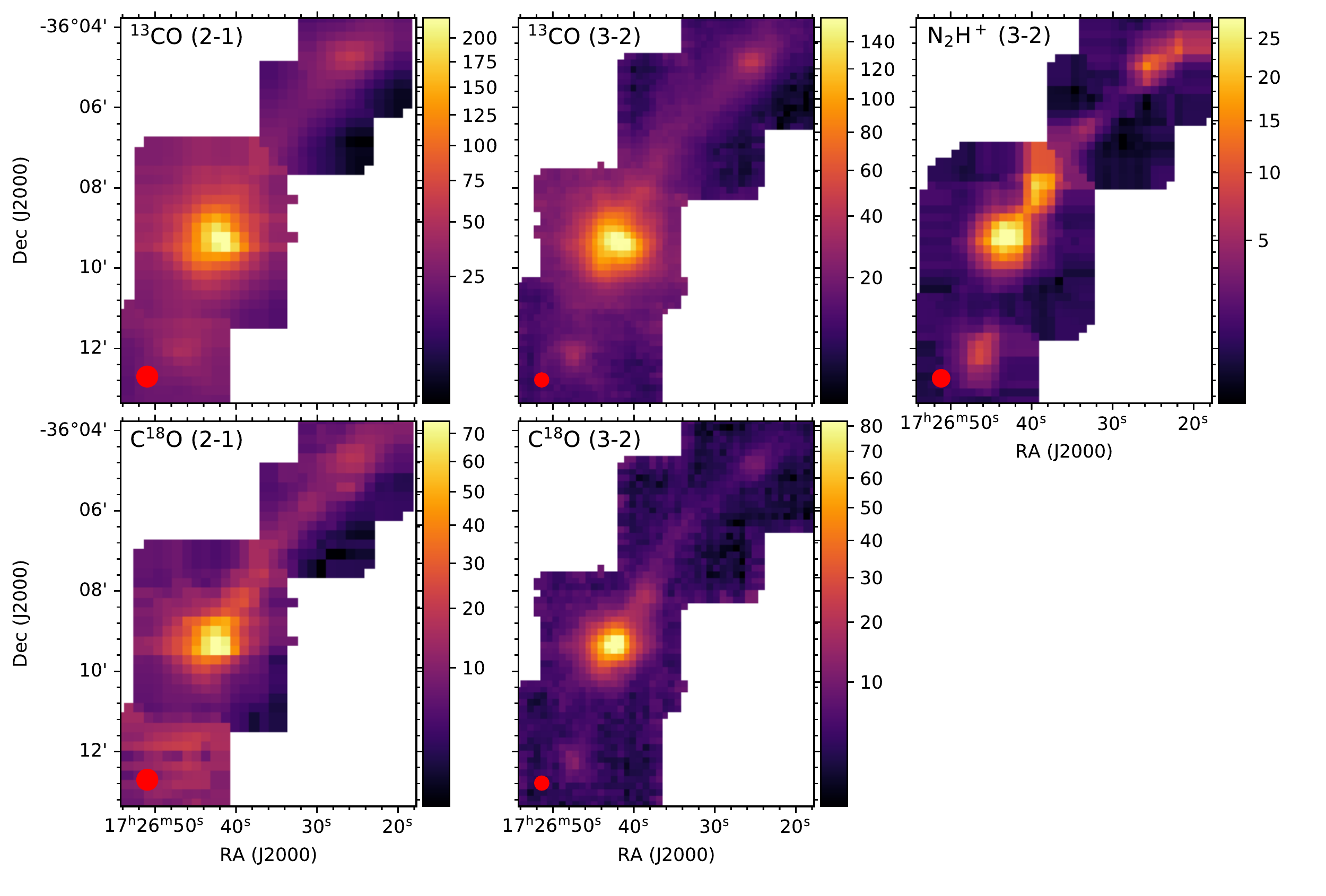} 
	\caption{Integrated intensity of the lines $^{13}$CO (2--1), $^{13}$CO (3--2), C$^{18}$O (2--1),	C$^{18}$O (3--2) and N$_2$H$^+$ (3--2). The red circles in the lower left corners indicate the beam size (HPBW).}
	\label{mom_0}
\end{figure*}

In Fig.~\ref{mom_0} we present maps of the integrated intensity (0th moment, M$_0$) in the lines $^{13}$CO(2--1), $^{13}$CO(3--2), C$^{18}$O(2--1), C$^{18}$O(3--2) and N$_2$H$^+$(3--2). The filamentary structure is visible in all maps.

\begin{figure}
	\includegraphics[width=1\linewidth]{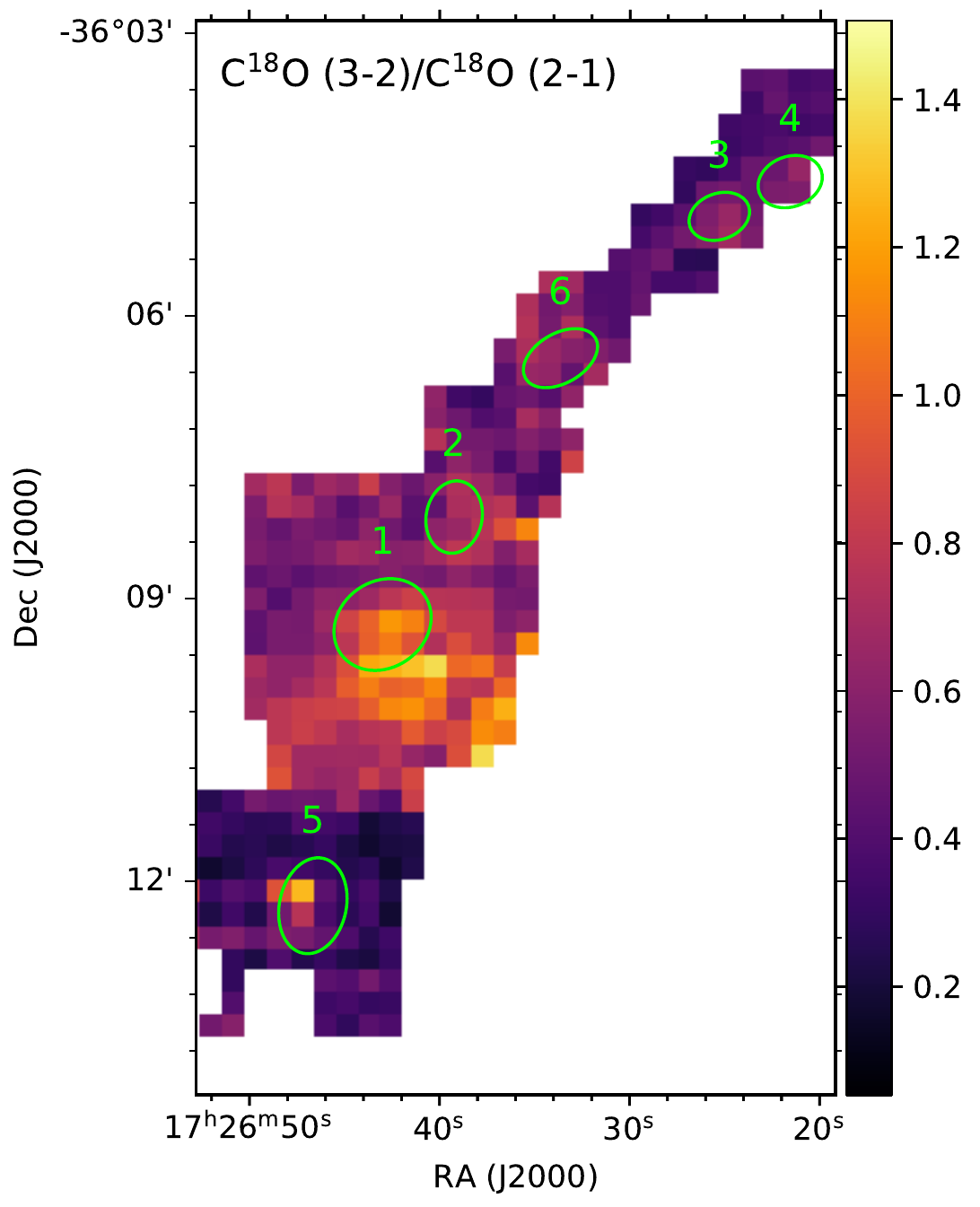} 
	\caption{Integrated intensity ratio of the lines C$^{18}$O (3--2)/C$^{18}$O (2--1). The ellipses show clumps detected by GaussClump using the dense gas tracer N$_2$H$^+$ (Sect. \ref{clump_sect}).}
	\label{ratio}
\end{figure}

Fig.~\ref{ratio} shows the map of the integrated intensity ratio C$^{18}$O(3--2)/C$^{18}$O(2--1). To construct this map the C$^{18}$O(3--2) map was smoothed to the same angular resolution as the  C$^{18}$O(2--1) map. This ratio may serve as an indicator of the physical conditions in the emission regions. The ellipses show clumps detected by GaussClump using a dense gas tracer N$_2$H$^+$ (Sect.~\ref{clump_sect}).

\subsection{Column density}\label{sect_colden}

In order to characterize the general distribution of the CO and N$_2$H$^+$ molecules we constructed maps of their column densities. 
The column densities of the C$^{18}$O and N$_2$H$^+$  molecules were calculated using the method presented in \cite{Mangum2015} on the basis of the C$^{18}$O(2--1) and N$_2$H$^+$(3--2) data. The first line is a tracer of the total molecular gas column density, while the second one is known to depict denser molecular gas compared to the C$^{18}$O data. We assume a low optical depth in the lines and the local thermodynamic equilibrium (LTE) conditions: 

\begin{eqnarray}\label{eq}
	N = \frac{3 h}{8 \pi^3 S \mu^2}  \frac{Q_{\rm rot}}{g_J g_K g_I} \frac{\text{exp}(\frac{E_u}{k T_{\rm ex}})}{\text{exp}(\frac{h \nu}{k T_{\rm ex}}) -1} \frac{\int T_{\rm B} dv }{J_\nu(T_{\rm ex}) - J_\nu(T_{\rm bg})}, \\
	J_\nu(T) = \frac{h \nu / k}{\text{exp}{\frac{h \nu}{k T}} - 1},
\end{eqnarray} 
where $h$ is the Planck constant, $S = J_u/(2J_u+1)$ is the line strength,  $\mu$ is the dipole moment, $Q_{\rm rot}$ is the partition function, $g_J$ = 2J + 1 is the rotational degeneracy, $g_K$ is $K$-degeneracy, $g_I$ is the degeneracy of the nuclear spin, $J_\nu(T)$ is the equivalent Rayleigh-Jeans temperature, $T_{\rm ex}$ is the excitation temperature, $T_{\rm bg}$ is the background temperature and the term $\int T_{\rm R} dv$ is the integrated intensity of the line. For C$^{18}$O and N$_2$H$^+$ molecules, g$_K$ = g$_I$ = 1 and $Q_{\rm rot}\simeq kT/hB + 1/3$, where $B$ is the rotational constant. %Dipole moments and rotational constants for molecules are taken from the JPL catalog\footnote{ https://spec.jpl.nasa.gov}. 
We assume that the molecules are thermalized with the excitation temperature being equal to the gas kinetic temperature, i. e., $T_{\rm ex} = T_{\rm kin}$, and that this equals the dust temperature, $T_{\rm dust}$. %, as derived in \cite{2019A&A...621A.130L}. 
We estimate the dust temperature from the \textit{Herschel} Hi-GAL data at 160-–500~$\mu$m using the approach described in \cite{2015MNRAS.447.2307M}. It is based on fitting the data with a modified black body spectrum. $T_{\rm dust}$ and $N$(H$_2$) are free parameters.
%Thermal emission of cold dust lies in the far infrared range, and its analysis can be used to obtain physical parameters, for example, the temperature of a dust $T_{\rm dust}$ and column density of of molecular hydrogen $N$(H$_2$) \citep{2011A&A...535A.128B, 2013A&A...551A..98L}. For this, the spectral distribution of the dust radiation energy was simulated according to the Herschel data (160–500 $\mu$m). For 4 points, a black body model was inscribed with free parameters $T_{\rm dust}$ and $N$(H$_2$) \citep{2015MNRAS.447.2307M}. 
As a result, maps of the distribution of these parameters were obtained. The temperature  lies in the range of 12 -- 23~K, and rises towards the IRAS 17233–3606 source. However, in the direction of the IRAS source itself, it was not possible to inscribe such a model, and a temperature of 25 K was adopted there, as in \cite{2019A&A...621A.130L}. In general, the temperature values are very close to the map presented in \citep{2019A&A...621A.130L}, so we do not show them here. Based on the 20\% calibration uncertainty assigned to the Hi-GAL fluxes,  a median temperature uncertainty is 13\% -- 18\% \citep{2011A&A...535A.128B}.

\begin{figure*}
		\includegraphics[width=1\linewidth]{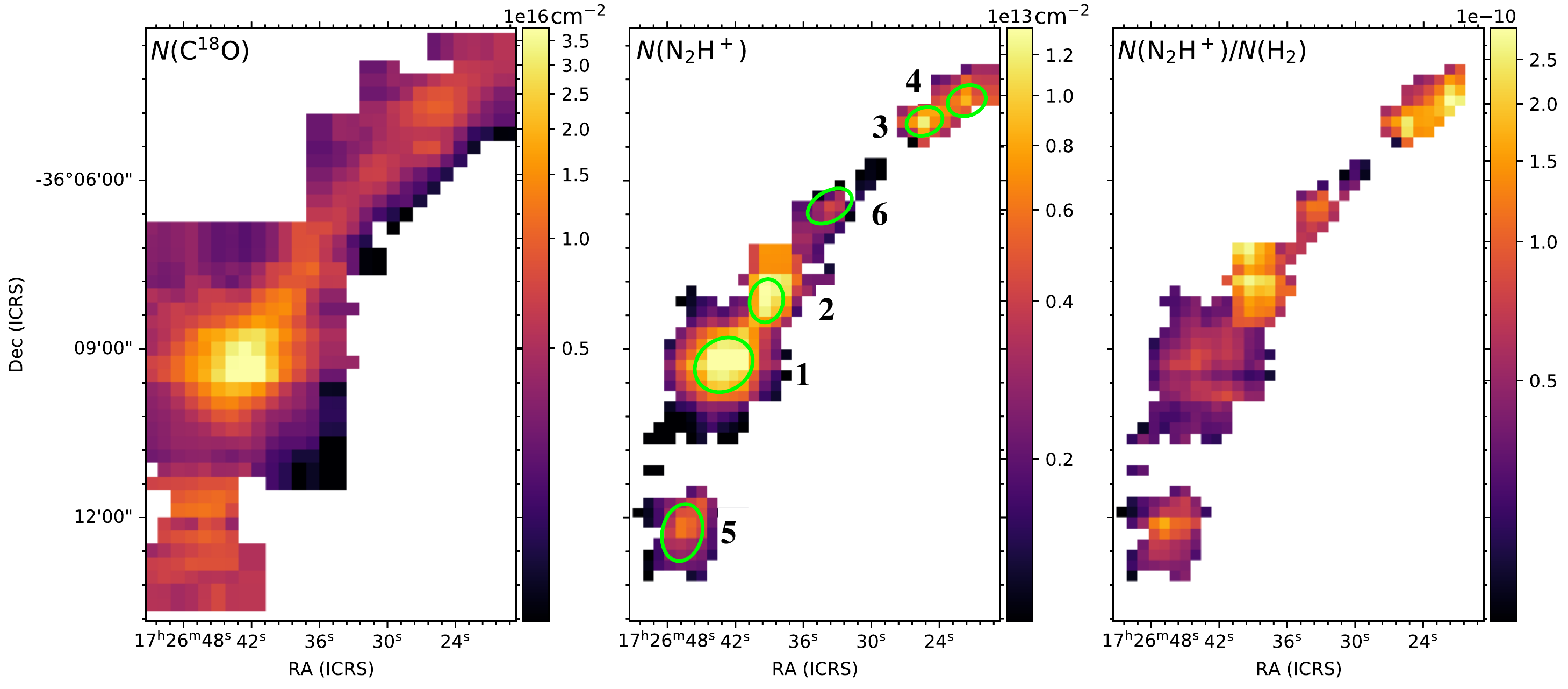} 
		\caption{The column density of C$^{18}$O (left), N$_2$H$^+$ (middle), and the relative abundance of N$_2$H$^+$ (right). The ellipses show clumps detected by GaussClump using a dense gas tracer N$_2$H$^+$ (Sect. \ref{clump_sect}).}
		\label{Nn2h}
\end{figure*}

Assumption of $T_{\rm kin}$ = $T_{\rm dust}$  is only accurate in dense regions ($n \ga 10^{5}$~cm$^{-3}$) shielded from the UV emission, however, the gas temperature may exceed the dust temperature near infrared sources \citep{Koumpia2015}. The C$^{18}$O column density is related to the hydrogen column density via a constant relative abundance C$^{18}$O/H$_2$ = 2.0$\cdot$10$^{-7}$ \citep{Liu2013} at the galactocentric distance of 7.4~kpc. The map of the hydrogen column density was obtained from the map of $N$(C$^{18}$O) using this relation. Although the hydrogen column density was estimated also from the Hi-GAL data, we use this map for consistency with the other works \citep[e.g][]{2019A&A...621A.130L}. Next, pixel-by-pixel division of the column density maps of N$_2$H$^+$ to H$_2$ was performed. The obtained maps of the distribution of the C$^{18}$O, N$_2$H$^+$ column density and N$_2$H$^+$ relative abundance are shown in Fig.~\ref{Nn2h}). On the map of the N$_2$H$^+$ column density dense clumps are marked by white ellipses (see Sect. \ref{clump_sect}). It can be seen that in the direction of the luminous IR source IRAS 17233--3606, the abundance of N$_2$H$^+$ decreases. Taking into account the uncertainties in the dust temperature and  integrated intensity of the C$^{18}$O and N$_2$H$^+$ lines, we find that the $N$(C$^{18}$O) uncertainty does not exceed 5\% and in the direction of dense clumps with a high signal-to-noise ratio it decreases to 3\%. Taking into account the negative feature in the C$^{18}$O (2--1) observations caused by the emission at the reference position (Sect.~\ref{sec:obs}), the column density is underestimated. Since the integrated intensity of this feature is $\sim 0.3$~K\,\kms, the corresponding column density is 1.5 $\cdot$ 10$^{14}$~cm$^{-2}$, while the median $N$(C$^{18}$O) is 6.7 $\cdot$ 10$^{15}$~cm$^{-2}$. Hence, the average underestimation does not exceed 2\%. The uncertainty in $N(\mathrm{N_2H^+})$ does not exceed 20\% in the direction of dense clumps, in other directions it reaches 50\%. 

By integrating the column density map of hydrogen we obtain the total mass of the filament as 1800 $\pm$ 50 M$_\odot$. The integration area is limited by the integrated intensity of the C$^{18}$O (2--1) line of 5~K\,km s$^{-1}$ (Fig.~\ref{g351_ik}). It roughly corresponds to the signal to noise ratio of about 5 for the most noisy C$^{18}$O (2--1) spectra.

The part of the filament we are studying has a length of 3.4 pc, and the mass to length ratio $M_{\rm line}$ = 529 M$_\odot$/pc. According to \cite{2019ApJ...877....1D}, the virial (or critical) line mass for a filament with non-thermal gas motions is calculated as
\begin{equation}
M_{\rm line,vir} =  \left[1 + \left(\frac{\sigma_{\rm NT}}{c_{\rm s}} \right)^2 \right] \times \left[16 {\rm M_\odot  pc^{-1}} \times \left( \frac{T}{10 \rm K} \right) \right],
\end{equation}
where $c_\mathrm{s}$ is the sound speed, $T$ is kinetic temperature and $\sigma_{\mathrm{NT}}$ is the non-thermal velocity dispersion, which is defined by:
\begin{equation}
\sigma_{\rm NT} = \sqrt{\frac{\Delta V^2}{8 \ln 2} - \frac{k T}{30 m_H}},
\end{equation}
where $\Delta V$ is the C$^{18}$O line width, $k$ is the Boltzmann constant, $m_{\rm H}$ is the mass of a hydrogen atom. The averaged over the entire filament C$^{18}$O (2--1) line width  is 2.5~\kms, the average temperature is 18~K. 
For our filament the critical line mass is $M_{\rm line,vir}$ = 512 M$_\odot$/pc. 
Taking into account the uncertainties, the mass to length ratio $M_{\rm line}$ is practically equal to the critical value. However, under the conditions of a possible falling motion inside the filament the velocity dispersion shows larger values than under virial equilibrium, therefore the $M_{\rm line,vir}$ artificially increases  \citep{2011MNRAS.411...65B}. Hence, the presented estimate is the upper limit of the linear critical mass. On the other hand, the derived mass to length ratio is also an upper limit since it does not take into account a possible inclination of the filament. In any case the observations show that the process of fragmentation is going on.

\subsection{Clumps}\label{clump_sect}

The GaussClumps algorithm, first proposed in  \cite{Stutzki1990}, was used to identify molecular clumps. In the position-position-velocity data cube, the absolute maximum of the emission is allocated, then a three-dimensional Gaussian is fitted into the position of this maximum, which is subtracted from the original cube. After that, the next maximum is searched, followed by inscription and subtraction. This happens until the criterion for the completion of the algorithm is satisfied.

In the present work, emission of the dense gas indicator N$_2$H$^+$(3--2) was used to identify clumps. Six clumps were found in this way, the following algorithm completion parameters were used: FWHM emission patterns in pixels (FwhmBeam) = 2.25, FWHM velocity is 3.0 km/s. Clump dimensions are defined as widths at the level of the half  intensity ($\Theta_{FWHM}$). The visualization of clumps is shown in Fig.~\ref{Nn2h}. The clump parameters are presented in Table~\ref{clumps_par}.

We obtain the mass of the gas by integrating the $N_{\mathrm{H}_2}$ column density over the surface of the clump $dA$ and the virial parameter of clumps  $\alpha_{\rm vir}$=M$_{\rm vir}$/M according to the definition in \cite{Ryabukhina2018}. To determine the mass of the clumps, the dust temperature obtained from the \textit{Herschel} data was taken (Sect. \ref{sect_colden}). Integration was carried out over regions with sizes two times larger than in Fig. \ref{Nn2h} in order to trace most of the clump emission, since the figure shows the sizes at the half intensity level. The resulting masses are presented in Table \ref{clumps_par}, where $\alpha_{2000}$ and $\delta_{2000}$ are coordinates of the center of the clump, $D$ is the clump size, FWHM (C$^{18}$O) is the C$^{18}$O (2--1) line width  (used to determine the virial mass), $T$ is the average dust temperature according to \textit{Herschel} data, $M$ is the clump mass, $M_{\rm vir}$ is the virial mass, $\alpha_{\rm vir}$ is the virial parameter. 

According to \cite{Kauffmann2013}, clumps are unstable and star formation processes can start when $\alpha_{\rm vir} \lesssim \alpha_{\rm crit} \simeq 2$. According to the values we obtained, this criterion is met by all clumps except the sixth one.

\begin{figure*}
	\includegraphics[width=1\linewidth]{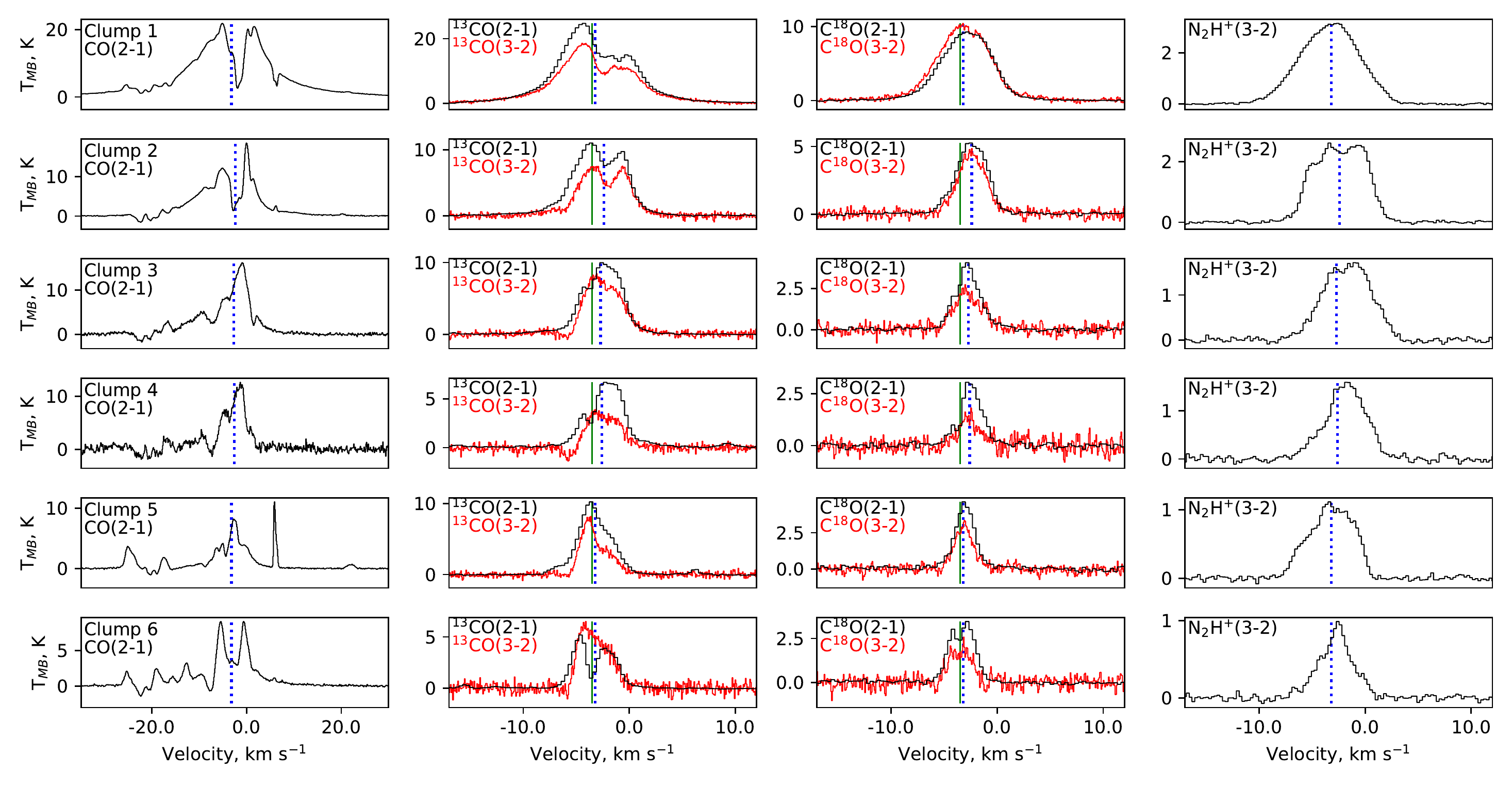} % Так вставляется рисунок
	\caption{Spectra of lines CO (2--1), $^{13}$CO (2--1), $^{13}$CO (3--2), C$^{18}$O (2--1), 	C$^{18}$O (3--2) and  N$_2$H$^+$ (3--2) in a clumps. The dashed lines indicate LSR velocity of each clump as obtained from the C$^{18}$O (3--2) line, the green solid lines indicate velocity of -- 3.5 \kms}
	\label{clumps_spectra}
\end{figure*}

Spectra averaged over clumps are shown in Fig.~\ref{clumps_spectra}. The blue dashed lines indicate the systemic velocity of each clump as $V_{\rm LSR}$ of the C$^{18}$O (3--2) line. The velocity values are given in Table \ref{clumps_par}. The green solid lines indicate the velocity of the emission  at the reference position in the $^{13}$CO and C$^{18}$O (2--1) lines. Optically thick CO (2--1) line shows a complex wide spectrum with dips. The maximum width (up to 10 km/s) is observed in the direction of the first clump, and such a width may indicate outflow in this direction. The spectra of $^{13}$CO J = 2--1, 3--2  are narrower.
The blue asymmetry of the $^{13}$CO (3--2) line seen in all clumps is indicative of infall \citep[e.g.][]{Snell1977}. The profiles of the $^{13}$CO (2--1) line can be influenced by the emission at the reference position. Nevertheless, they also show real self-absorption dips at least in the first and second clumps (at about --2~\kms).
The optically thin C$^{18}$O line has an even smaller width and smaller dips, and in N$_2$H$^+$ (3--2) toward the first clump there is no dip at all. 

It is worth mentioning the apparent velocity shift between N$_2$H$^+$ (3--2) and CO isotopologues in some clumps. We discuss this later.

According to Fig.~\ref{ratio}, the ratio of the integrated intensities C$^{18}$O(3--2)/C$^{18}$O(2--1) reaches a value $\sim 1.5$ toward the first clump, however, in Fig. \ref{clumps_spectra} the intensities are approximately equal. The reason is that in Fig. \ref{clumps_spectra} spectra averaged over clumps are presented; therefore, the high ratio, which appears in some pixels, is smoothed out.

\begin{table*}
	\caption{  {Clumps parameters}}
	\label{clumps_par}
	%\bigskip
	\begin{tabular}{ccccccccccc}
		\hline
		Clump & \specialcell{$\alpha_{2000}$,\\ h m s}  &  \specialcell{$\delta_{2000}$,\\ $\circ $ $ \prime$ $\prime\prime$ }  & \specialcell{D, \\ pc} & \specialcell{ FWHM C$^{18}$O (2--1), \\ \kms} & \specialcell{$V_{\rm LSR}$ C$^{18}$O (3--2) , \\ \kms} &  \specialcell{T, K \\Herschel} &   M, M$_\odot$  &  M$_{vir}$, M$_\odot$ & $\alpha_{vir}$ \\ \hline
		1 & 17 26 43 & -36 09 18 & 0.30 $\times$ 0.26 & 5.46 $\pm$ 0.04  & -- 3.15 $\pm$ 0.01 & 25 &	635 $\pm$ 25 &	889  & 1.4   \\
		2 & 17 26 39 & -36 08 09 & 0.17 $\times$ 0.22 & 3.39 $\pm$ 0.03 & -- 2.39 $\pm$  0.02 & 18 & 173 $\pm$ 13	& 235  & 1.35  \\
		3 & 17 26 25 & -36 04 58 &0.19 $\times$ 0.14 & 2.73 $\pm$ 0.05 & -- 2.76 $\pm$ 0.04 & 15 & 84 $\pm$ 9 &	130 & 1.55   \\
		4 & 17 26 21 & -36 04 36 & 0.20 $\times$ 0.15 & 2.32 $\pm$ 0.06 & -- 2.65 $\pm$ 0.06	&14 & 57 $\pm$ 8	&  100  & 1.75  \\
		5 & 17 26 46 &-36 12 17 & 0.20 $\times$ 0.29 & 2.17 $\pm$ 0.04  & -- 3.15	$\pm$ 0.02 & 17 & 160 $\pm$ 13	& 121 &  0.76     \\
		6 & 17 26 33 & -36 06 28 & 0.25 $\times$ 0.15 & 2.85 $\pm$ 0.07  &	-- 3.46 $\pm$ 0.04 & 14 & 77 $\pm$ 9 & 166 & 2.15    \\
		[1mm]
		\hline
	\end{tabular}
\end{table*}

\begin{figure}
	\includegraphics[width=1\linewidth]{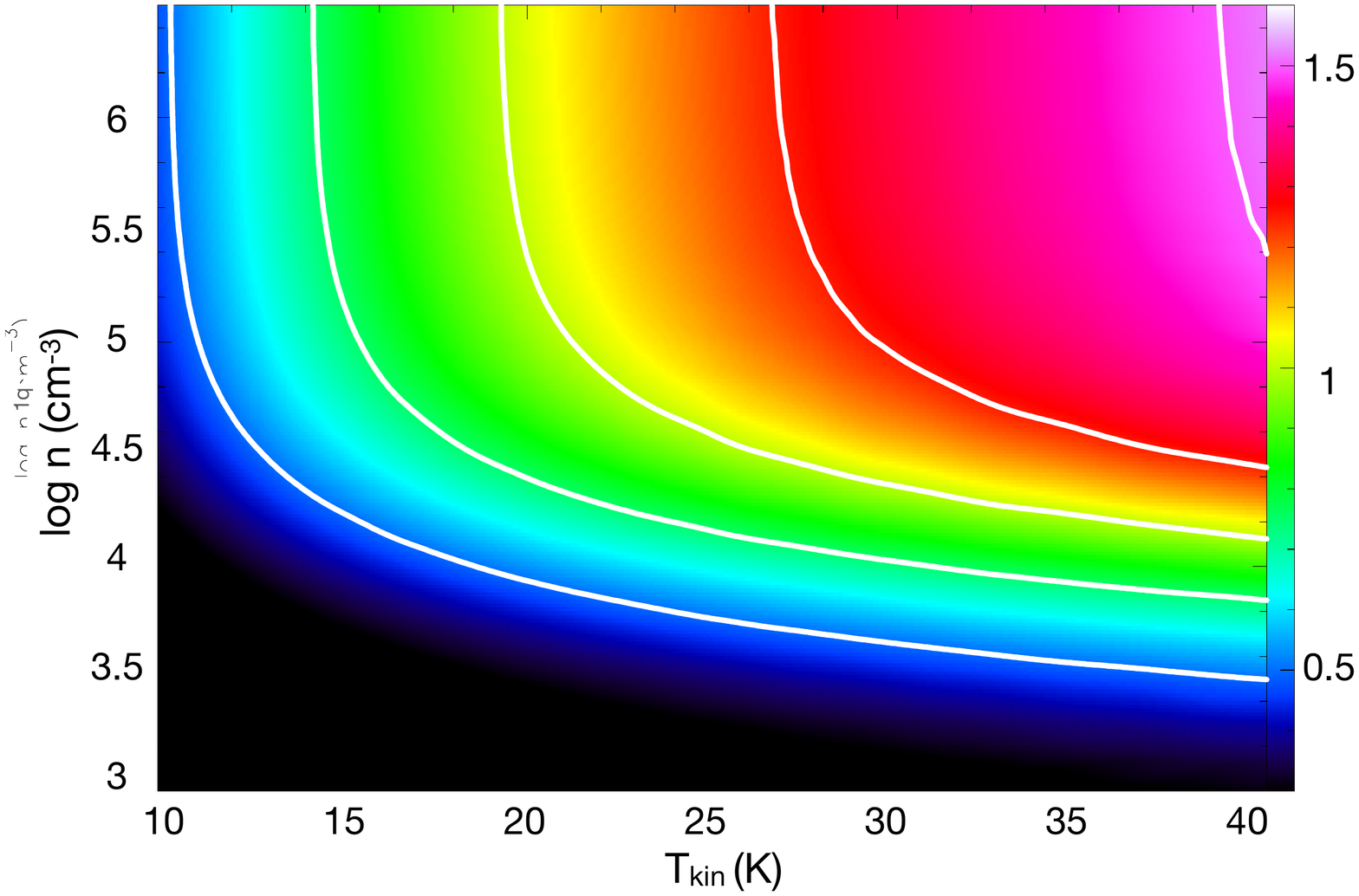} 
	\caption{Model dependence of the intensity ratio of C$^{18}$O (3--2)/C$^{18}$O (2--1) on the gas temperature and density. White contours correspond to ratios 0.5, 0.75, 1.0, 1.25, 1.5.}
	\label{radex}
\end{figure}

We modelled the C$^{18}$O (3-2)/(2-1) ratio with the Radex package (Fig. \ref{radex}) assuming the C$^{18}$O column density sufficiently low to ensure a low optical depth in the lines. By comparison with the observations this model places some constraints on the density and/or temperature of the clumps. For the first clump the observed ratio reaches a value $\sim 1.5$. This is inconsistent with the temperature of 25~K from the Herschel data (Table~\ref{clumps_par}). However for the average spectra the ratio is $\sim 1$, which implies density $\sim 4\times 10^4$~\cmc at $T_k = 25$~K. For the second clump the average ratio is also $\sim 1$. At the temperature of 18~K from Herschel this implies a very high density $\ga 10^6$~\cmc. However, our CH$_3$CCH data indicate a somewhat higher temperature (see below) which is consistent with the density similar to that in the first clump. For the other clumps the C$^{18}$O (3-2)/(2-1) ratio varies from $\sim 0.5$ to $\sim 0.75$ and their temperature according to Herschel is $\sim 15$~K. This implies densities $\sim 3\times 10^4$~\cmc. 

\begin{figure*}
	\includegraphics[width=1\linewidth]{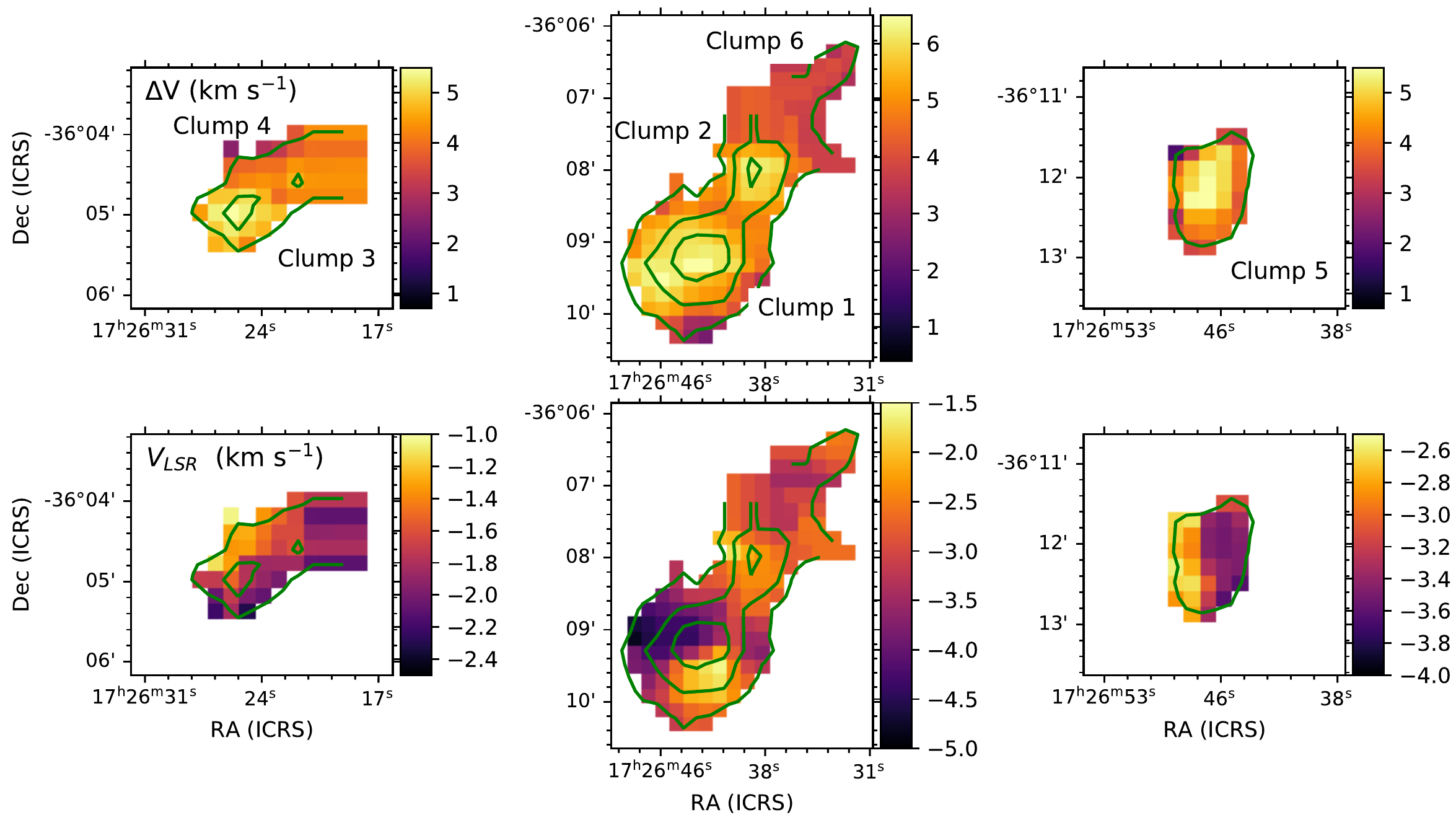} 
	\caption{Maps of the first moment (bottom) and line width (top) in the N$_2$H$^+$ (3--2) line toward various clumps. Green contours show the integrated intensity of the N$_2$H$^+$ line, the levels are at 3, 10, 20 K\,\kms}
	\label{fig:moments}
\end{figure*}

In order to better characterize kinematics of the clumps we present in Fig.~\ref{fig:moments} maps of the first moment (which correspond to the velocity) and the line width (obtained from fitting by a single Gaussian) in the N$_2$H$^+$(3--2) line toward all six clumps. In most clumps there are apparent velocity gradients. The line width peaks in the center of the clumps. We discuss these maps in Sect.~\ref{sec:disc}.

\begin{figure}
	\center{\includegraphics[width=1\linewidth]{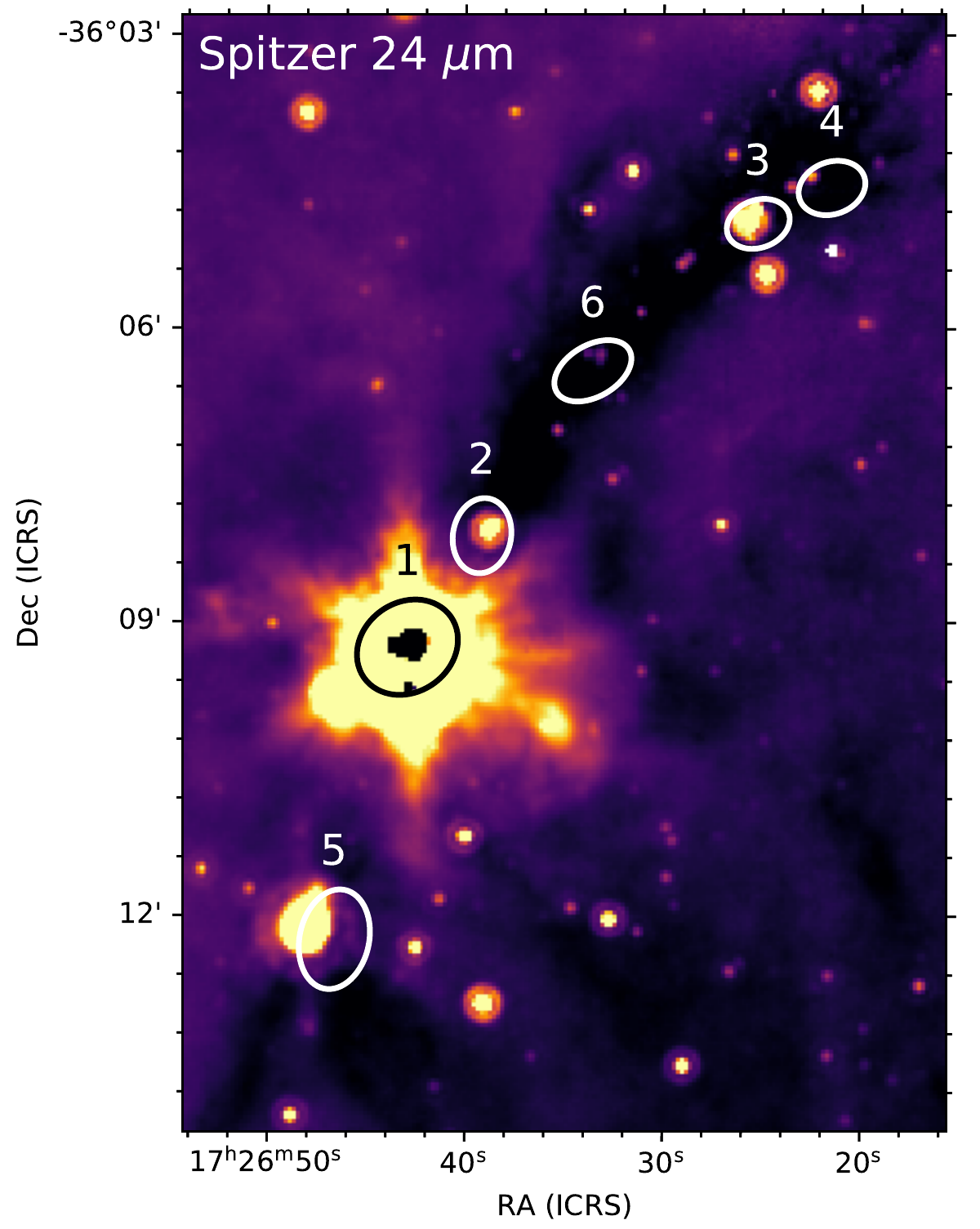}}
	\caption{The map of the region G351.78--0.54 in the infrared range at $\lambda$ = 24~$\mu$m (Spitzer). Ellipses show clumps detected by GaussClump using a dense gas tracer N$_2$H$^+$} 
	\label{g351_24m}
\end{figure}

Fig. \ref{g351_24m} shows the map of the region G351.78--0.54 in the mid-infrared range at $\lambda$ = 24~$\mu$m (Spitzer) with clumps, obtained in this work. There is a chain of mid-IR sources along the filament, but in clumps 1--3 and 5 they are the most luminous.

\subsection{Kinetic temperature from CH$_3$CCH}

\begin{figure}
	\includegraphics[width=1\linewidth]{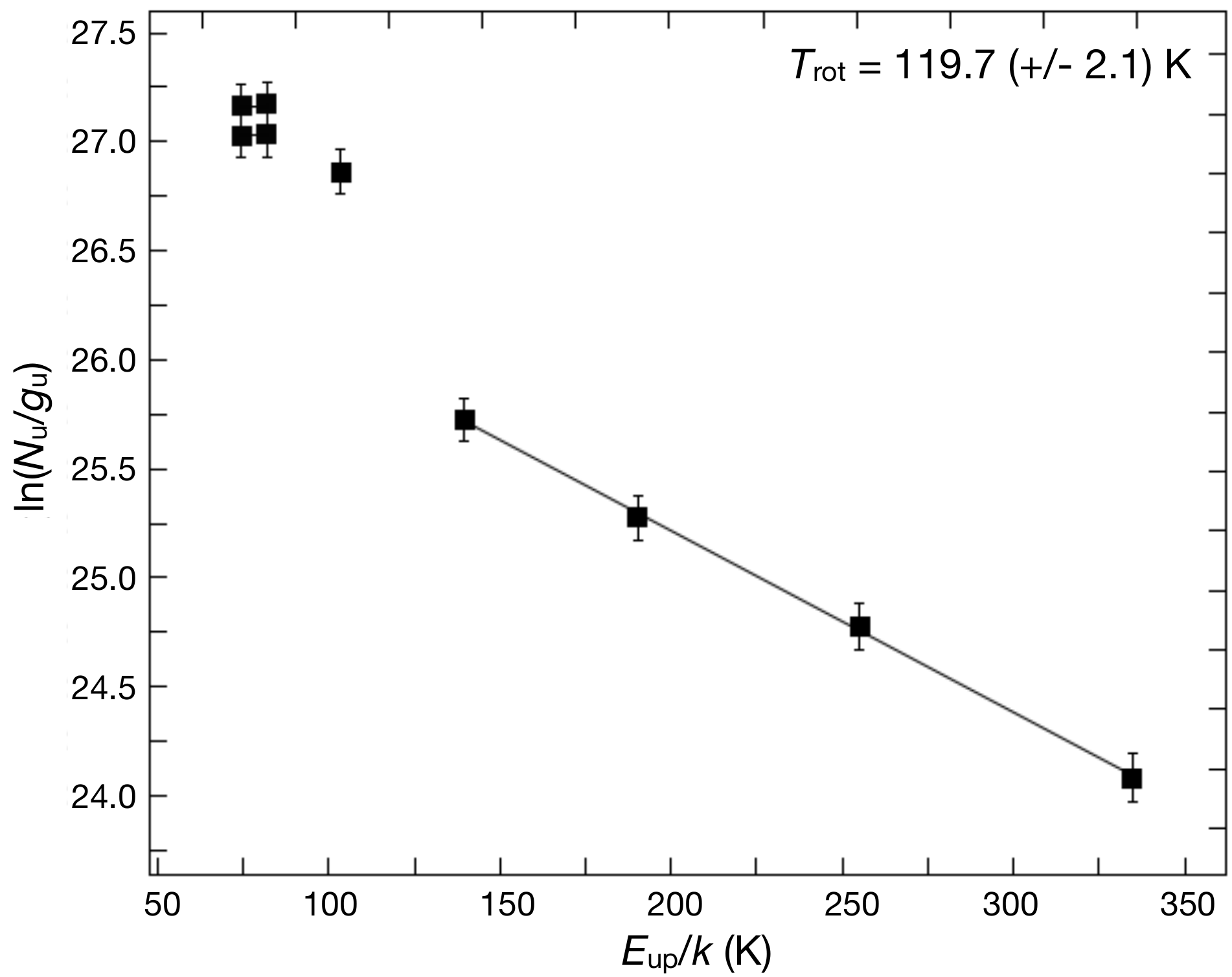} % Так вставляется рисунок
	\caption{The rotation diagram for the CH$_3$CCH transitions detected toward IRAS~17233--3606}
	\label{temp_ch3cch}
\end{figure}

\begin{figure}
	\includegraphics[width=1\linewidth]{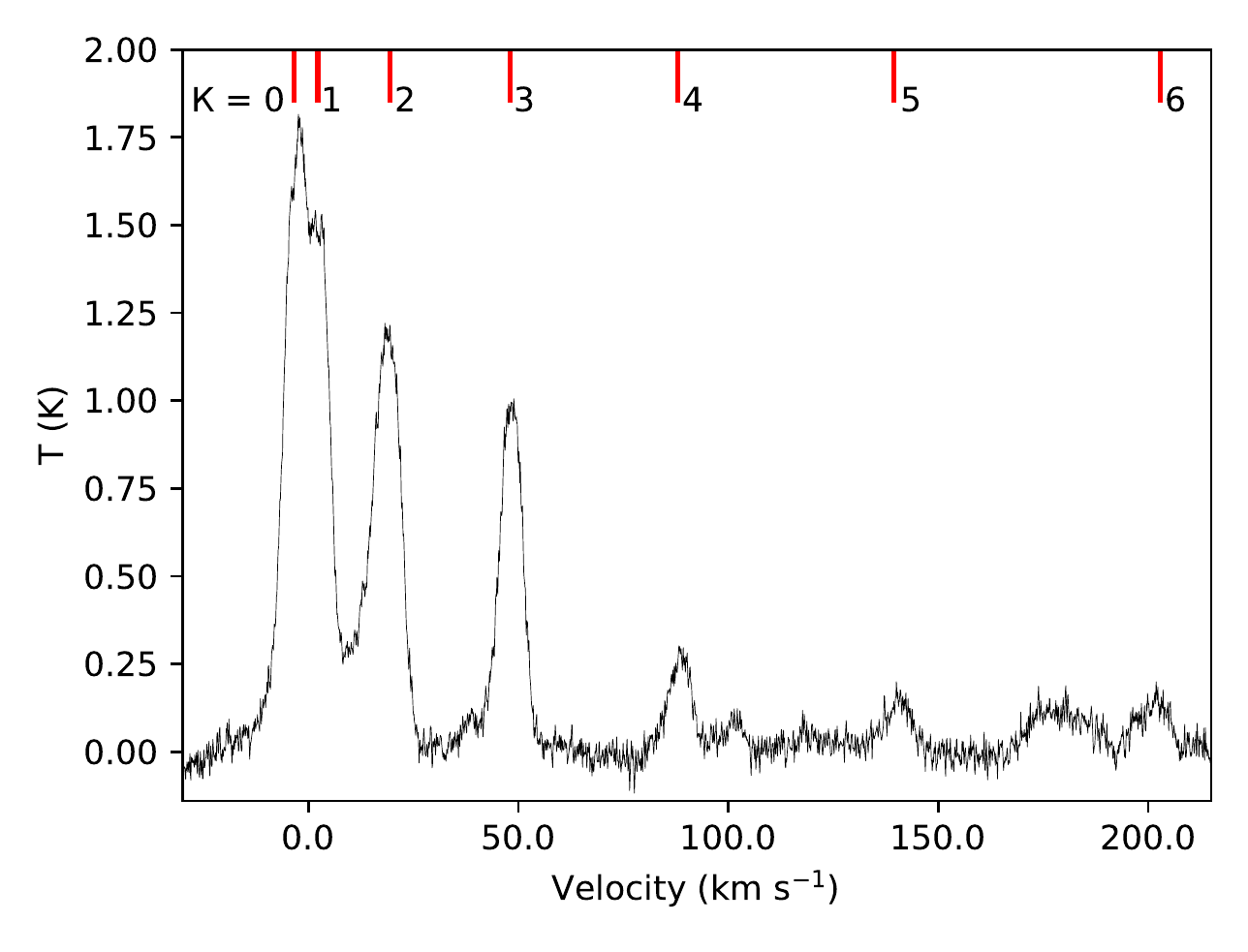} % Так вставляется рисунок
	\caption{The CH$_3$CCH spectrum toward IRAS~17233--3606}
	\label{spec_ch3cch}
\end{figure}

The kinetic temperature of the gas was estimated from observations of the CH$_3$CCH (13--12) line  using the ``rotation diagram" or ``population diagram" method \citep{Goldsmith1999}. The methylacetylene molecule is a type of symmetric top and is a reliable indicator of kinetic temperature even at rather low densities  \citep{Bergin1994}. These calculations were performed using the Cassis program. For several lines of the same rotational transition with different values of the $K$ projection for a molecule of the symmetric top type, under the LTE conditions the following relation holds:
\begin{equation*}
\ln\frac{N_{\rm u}}{g_{\rm u}} = \ln\frac{N_{\rm tot}}{Q(T_{\rm rot})} - \frac{E_{\rm U}}{k T_{\rm rot}},
\end{equation*}
where $N_\mathrm{u}$ is the  population of each level, $g_u$ is statistical weight, $N_{\mathrm{tot}}$ is total column density. Up to 7 lines of the CH$_3$CCH (13--12) transition are detected (see Table~\ref{table:lines}), for which we build the dependence of $\ln\frac{N_{\mathrm{u}}}{g_{\mathrm{u}}}$ on $\frac{E_{\rm U}}{k}$. The kinetic temperature $T$ is found as the reciprocal of the slope of a straight line approximated by this dependence. A total of six positions were observed along the filament, however, a sufficiently strong emission is detected only toward the first and second clumps. An example of the $\ln \frac{N_{\mathrm{u}}}{g_{\mathrm{u}}}$ dependence for the first spectrum in the IRAS 17233--3606 direction is shown in the Fig. \ref{temp_ch3cch}. In this direction, 7 CH$_3$CCH (13--12) lines are detected (Fig. \ref{spec_ch3cch}), however, the first two lines J = 13$_0$ -- 12$_0$ and 13$_1$ -- 12$_1$ are blended and it is impossible to reliably determine the integrated intensities for them. The temperature derived from the higher excitation lines from 13$_3$--12$_3$ to 13$_6$--12$_6$ is 119.7 $\pm$ 2.1 K. In the direction of the second clump, lines from 13$_0$--12$_0$ to 13$_3$--12$_3$ are observed, and the temperature according to the rotation diagram is 26 $\pm$ 4.5 K.

\subsection{HNCO toward IRAS 17233--3606}

\begin{figure}
	\includegraphics[width=\linewidth]{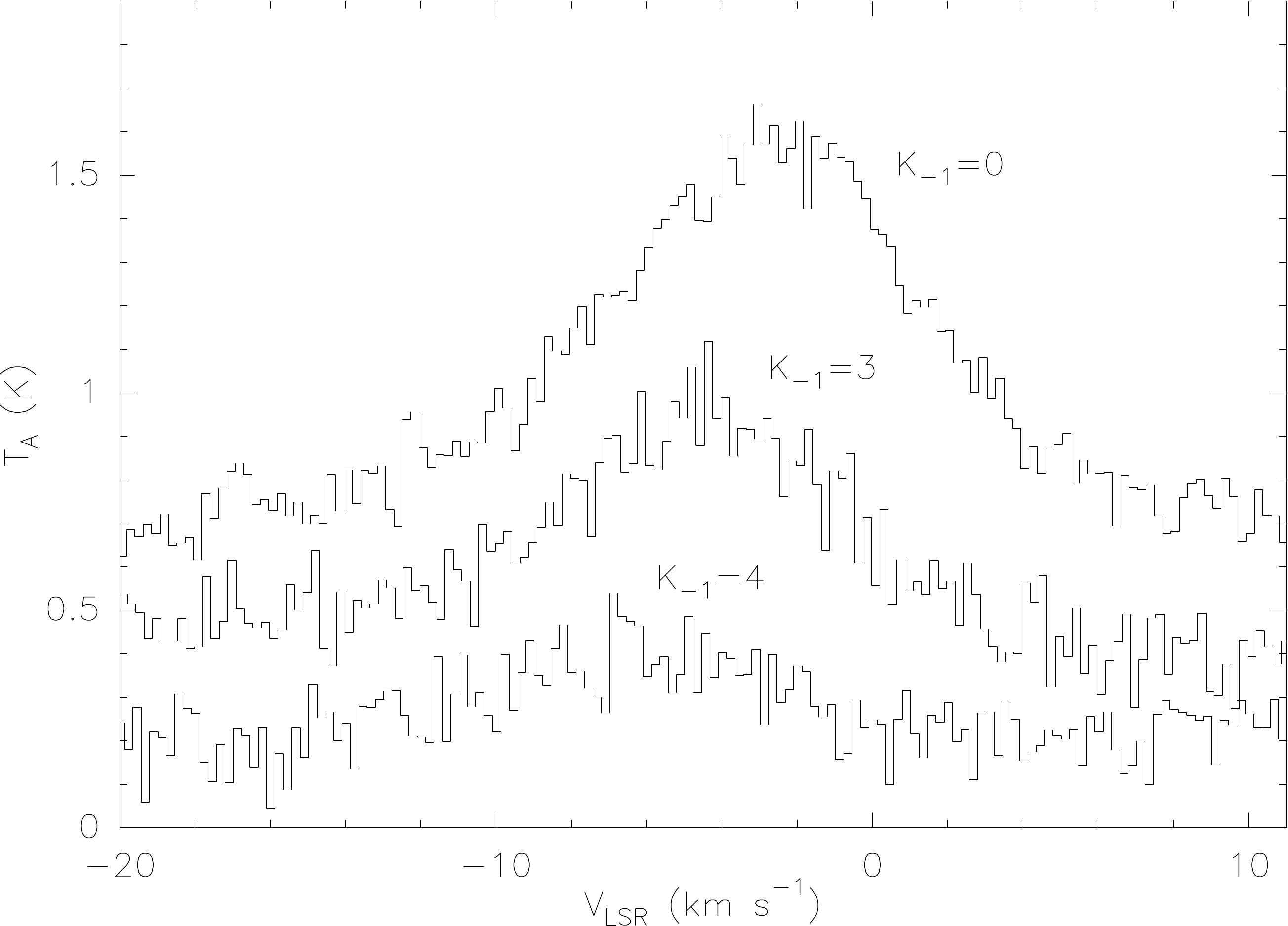} 
	\caption{Examples of the HNCO $J=15-14$ spectra measured toward IRAS 17233--3606. No baseline correction was applied. The spectra are shifted along the ordinate axis for clarity.}
	\label{fig:hnco-spectra}
\end{figure}

HNCO is a valuable probe of high mass star-forming regions \citep{Zinchenko2000}. Our data set includes several HNCO lines, which belong to the $J=10-9$ and $J=15-14$ transitions with the excitation energies up to $\sim 800$~K above the ground level (Table~\ref{table:lines}). The HNCO emission is detected exclusively toward the IR source IRAS~17233--3606. Examples of the detected lines are given in Fig.~\ref{fig:hnco-spectra}. The size of the emission region in the $J_{K_{-1}}=10_0-9_0$ line is comparable to the beam size. A 2D Gaussian fit gives the size $\approx 41^{\prime\prime}\times 34^{\prime\prime}$. \cite{Leurini2011} measured with the Submillimeter Array (SMA) the deconvolved size in this line $2\farcs5\times 2\farcs1$. The emission in the $J_{K_{-1}}=10_3-9_3$ lines was point-like with their beam ($5\farcs4\times 1\farcs9$). A comparison of our HNCO line intensities with those measured by \cite{Leurini2011} shows a significant flux loss (by a factor of 3) in the $J_{K_{-1}}=10_0-9_0$ line observations with the SMA. At the same time there is no flux loss for the $J_{K_{-1}}=10_3-9_3$ line. It shows that the emission in the latter line is really point-like while the $J_{K_{-1}}=10_0-9_0$ emission contains an extended component resolved out with the SMA. 

%\begin{figure}
%	\includegraphics[width=\linewidth]{hnco10-9_map+wings_gr} 
%	\caption{Map of the integrated intensity in the HNCO $J_{K_{-1}}=10_0-9_0$ transition (color scale). The blue and red contours show the emission in the blue and red wings of this line, respectively. The beam at the half intensity level (HPBW) is shown in the lower left corner.}
%	\label{fig:hnco-map}
%\end{figure}

The $J_{K_{-1}}=10_0-9_0$ and $J_{K_{-1}}=15_0-14_0$ emission spectra show broad wings, which most probably arise in the outflow. It is known that HNCO is an outflow tracer \citep{Zinchenko2000}. %Contours of the blue-shifted and red-shifted HNCO wing emission are plotted in Fig.~\ref{fig:hnco-map}. 
The orientation of the HNCO outflow is in a good agreement with the observations of the outflowing gas in the lines of other molecules \citep{Leurini2008,Klaassen2015}.

\begin{figure}
	\includegraphics[width=\linewidth]{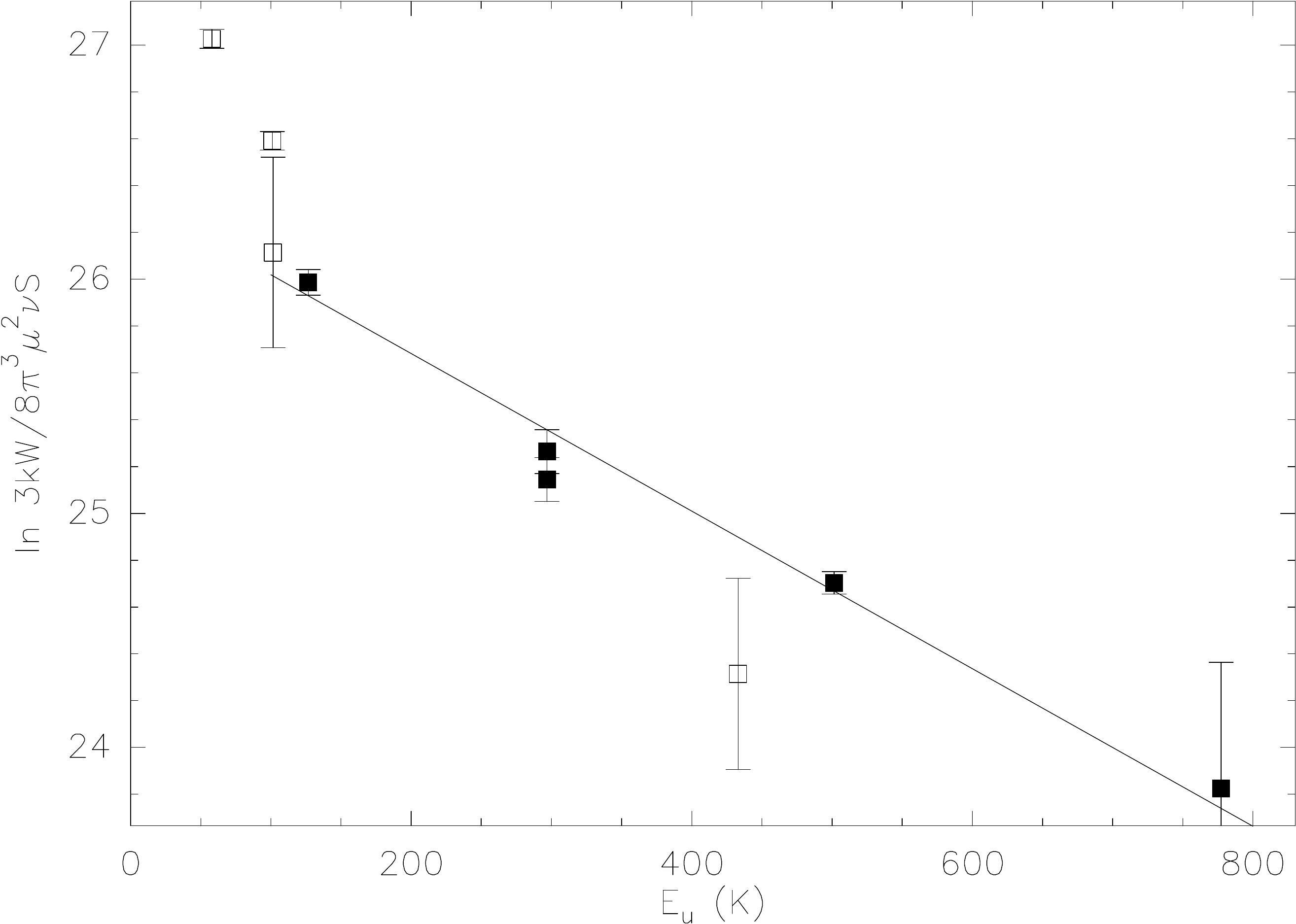} 
	\caption{The rotation diagram for the HNCO transitions detected toward IRAS~17233--3606. The open symbols correspond to the $J=10-9$ transitions and the filled symbols correspond to the $J=15-14$ transitions. The solid line represents a least squares fit to the $J=15-14$ data. Its slope corresponds to the rotational temperature of $T_{rot}\approx 300$~K.}
	\label{fig:hnco-rd}
\end{figure}

The rotation diagram for the HNCO lines detected toward IRAS~17233--3606 is presented in Fig.~\ref{fig:hnco-rd}. The $J=15-14$ transitions are well fitted by a single component with the rotational temperature of $T_{rot}= 297\pm 8$~K. The excitation energy of the upper levels of these transitions exceeds 100~K. The $J=10-9$ transitions with lower excitation temperatures clearly indicate a lower rotational temperature. However its a more or less reliable estimate from these data looks difficult. A rough value is between 50 and 100~K. The point corresponding to the $J_{K_{-1}}=10_3-9_3$ transitions with the excitation energy of the upper level about 430~K lies somewhat lower than the fit to the $J=15-14$ data. This can be explained by the source compactness since no correction for the beam size was applied. Taking into account the difference in the beam sizes for the $J=10-9$ and $J=15-14$ transitions, this point is in a good agreement with the least squares fit mentioned above.

\begin{figure}
	\includegraphics[width=\linewidth]{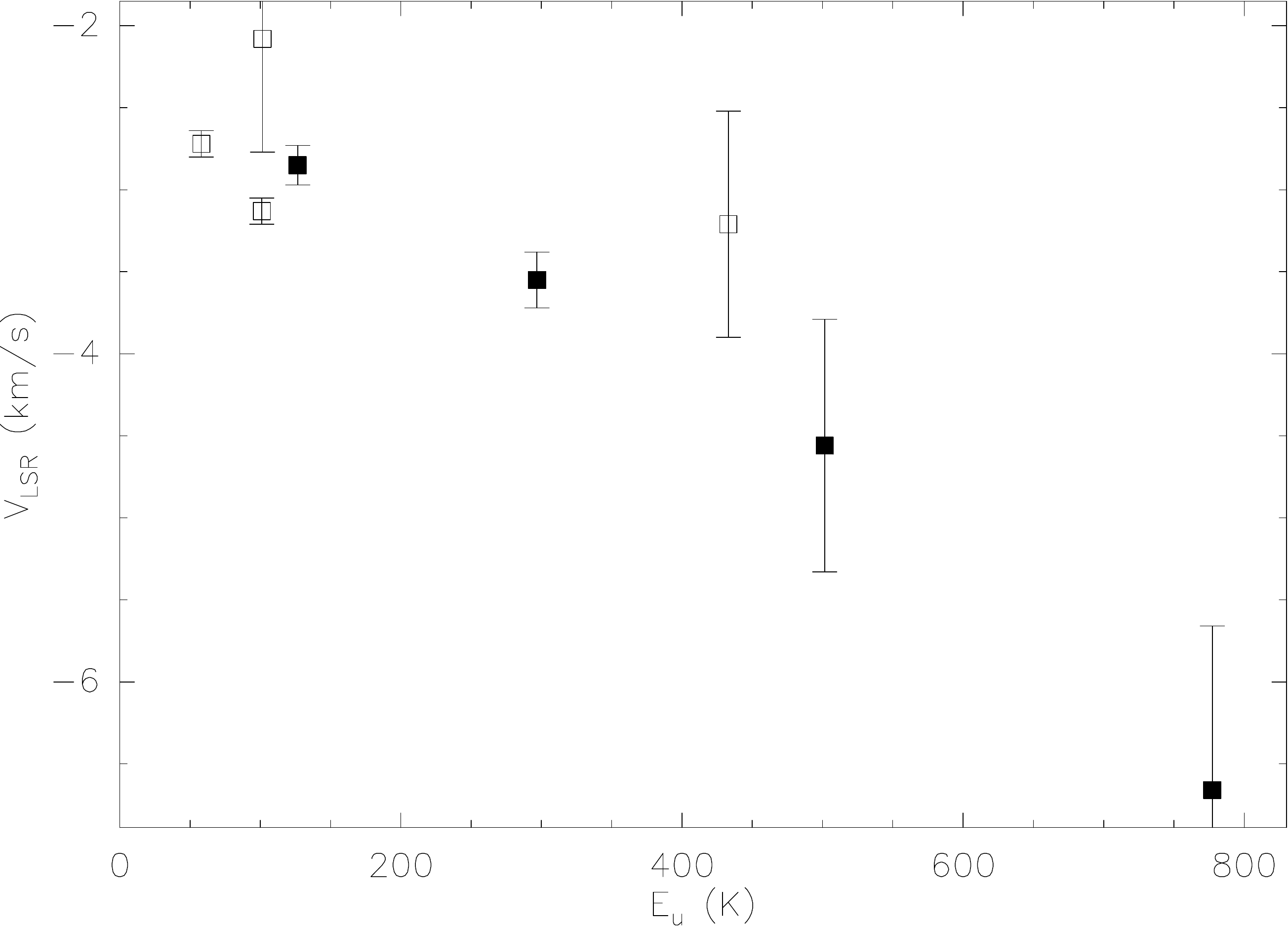} 
	\caption{The LSR velocities measured in the HNCO lines detected toward IRAS~17233--3606 in dependence on the excitation energy of the upper level. The symbols are the same as in Fig.~\ref{fig:hnco-rd}.}
	\label{fig:hnco-vel}
\end{figure}

\begin{figure}
	\includegraphics[width=\linewidth]{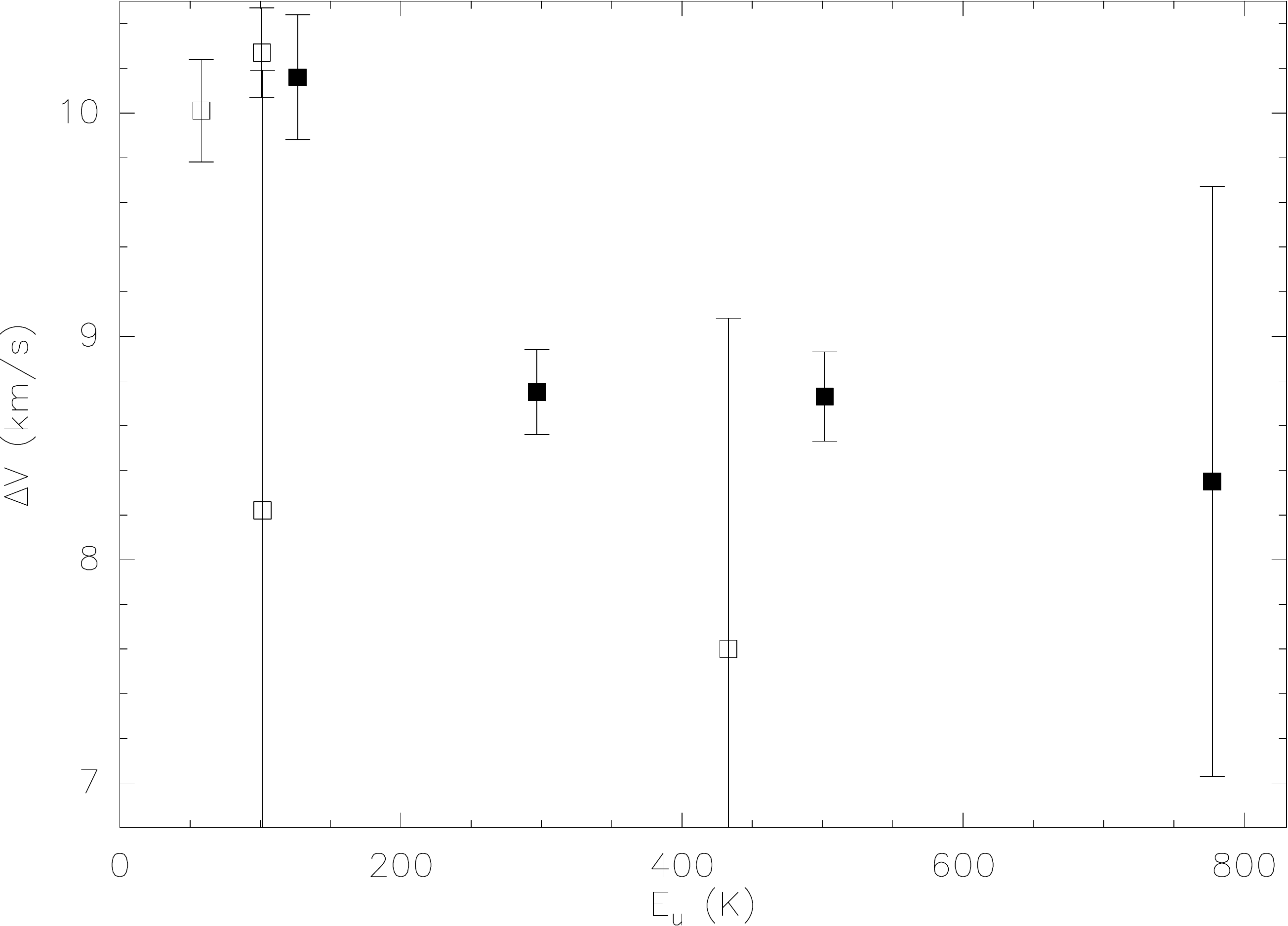} 
	\caption{Widths the HNCO lines detected toward IRAS~17233--3606 in dependence on the excitation energy of the upper level. The symbols are the same as in Fig.~\ref{fig:hnco-rd}.}
	\label{fig:hnco-width}
\end{figure}

An interesting feature in the HNCO data is the apparent dependence of the central velocity (as derived from Gaussian fitting) on the excitation energy of the transition (Fig.~\ref{fig:hnco-vel}). This shift greatly exceeds the measurement uncertainties and cannot be explained by instrumental effects since some of the high-excitation HNCO lines are close in frequency to other strong lines in this source, in particular C$^{18}$O, which are observed at a ``normal" velocity. We also cannot explain this picture by a possible misidentification of the high-excitation HNCO lines. No other reasonable identification of these lines could be found. Therefore, this dependence reflects apparently the internal kinematics of the source. It is worth noting that \cite{Leurini2011} also measured a significant difference in the velocities of the $J_{K_{-1}}=10_3-9_3$ and $10_0-9_0$ transitions in the same sense as in our data. This dependence makes the rotational diagram analysis questionable, since the emission in different transitions comes apparently from different regions. At the same time the widths of the higher excitation HNCO lines ($E_u\ga 200$~K) do not show any dependence on the excitation energy, while the lower excitation lines are somewhat broader (Fig.~\ref{fig:hnco-width}). These broader line widths include apparently the contribution from the line wings arising in the outflow.

\section{Discussion} \label{sec:disc}

In \cite{2011A&A...533A..85L}, a search for clumps by the ClumpFind method was also performed in the maps of the 870 $\mu$m continuum and their parameters were found. The position of our first clump corresponds to the first clump from \cite{2011A&A...533A..85L}, the second corresponds to the second, third to fifth, fourth to sixth, fifth to third, sixth to seventh. The numbering of our clumps is based on the intensity of the emission peak, which decreases with increasing the number. The clumps we found are relatively large (for example, 43'' $\times$ 38'' in \cite{2011A&A...533A..85L} and 64'' $\times$ 54'' for our first clump). The intensities and profiles of the  $^{13}$CO (2--1) and C$^{18}$O (2--1) lines presented here and in \cite{2011A&A...533A..85L} are similar in general. Some differences can be related to the fact that we present spectra averaged over the clumps, while in \cite{2011A&A...533A..85L} spectra at single positions are given. In addition, the spectra, especially $^{13}$CO (2--1) can be influenced by the emission at the reference position (Sect.~\ref{sec:obs}).

The filament contains sources at different stages of evolution. The central source  IRAS 17233--3606 is considered the most evolved and probably represents a region of massive star formation  \citep{Yu2018}. Three protostellar objects (AGAL351.774-00.537, 351.784-00.514, 351.804-00.449)  are distributed along the filament. AGAL351.774-00.537 corresponds to the first clump, AGAL351.784-00.514 corresponds to the second clump, AGAL351.804-00.449 corresponds to the third clump. In the southern part of the filament there is a region of ionized hydrogen AGAL351.744-00.577, which is associated with our fifth clump \citep{Contreras2013}. It is worth noting that all our clumps are associated with the mid-IR sources (Fig.~\ref{g351_24m}), which confirms their protostellar nature. 

In our analysis the main filamentary body  has a mass of $\sim 1800$~M$_\odot$. This value is somewhat higher than obtained by \cite{2019A&A...621A.130L} using dust temperature from \textit{Herschel} and C$^{18}$O (2--1) molecular data ($\sim 1200$~M$_\odot$). However, the estimate of the filament mass from the dust emission using the Hi-GAL column density map is $\sim$\,1870 M$_\odot$ disregarding the region near the first clump, which is $\sim$\,190 M$_\odot$ \citep{2019A&A...621A.130L}. Our filament mass is closer to the latter value. 
As shown in Sect.~\ref{sect_colden} 
 the mass to length ratio $M_{line}$ is practically equal to the critical value, although both values represent upper limits. Nevertheless, the presence of several dense clumps along the filament shows that the process of fragmentation is going on.

The average clump densities estimated from the $J = 2-1$ and $J = 3-2$ C$^{18}$O line intensity ratio are $\sim (3-4)\times 10^4$~\cmc\ (Sect.~\ref{clump_sect}). Another estimate can be obtained from comparison of our N$_2$H$^+$(3--2) data with the N$_2$H$^+$ (1--0) observations in \cite{2011A&A...533A..85L}.

In \cite{2011A&A...533A..85L} spectra of the N$_2$H$^+$ (1--0) are presented in the direction of the first, second and fifth clumps (which correspond to our first, second and third one). In the first clump, the spectrum has the antenna temperature $\sim$ 2 K, and very broad lines: 7 hyperfine components merge into 2, and in the direction of the second and fifth clumps, the temperature is 1.4 and 1.2 K, respectively, and the lines merge into three. According to the MALT90 survey \cite{2016PASA...33...30R}, the N$_2$H$^+$ (1--0) antenna temperature of the first clump is 2.2 $\pm$ 0.04 K, the line width is 5.03 $\pm$ 0.08 km/s. This width was obtained by approximating the line with three Gaussian functions.  
The N$_2$H$^+$ (3--2) line splits into 38 hyperfine components closely spaced in frequency \citep{2009A&A...494..719P}. Due to turbulent line broadening, the components merge into one. We fit the observed profiles by a set of these components assuming a low optical depth in the line and equal excitation temperatures for all of them. For example, for the first clump, the line can be approximated by a single Gaussian function with an amplitude of 2.23 $\pm$ 0.01 K and a line width of 6.5 $\pm$ 0.03~\kms, but the multi-component fit gives the line width of 6.33 $\pm$ 0.02~\kms, and the intensity of the brightest component of 0.401 $\pm$ 0.001~K. The parameters of the hyperfine structure of the N$_2$H$^+$ (3--2) line for the all clumps are shown in Table~\ref{n2h_hpf_par}, where $T$ is the maximum intensity for the $F_1F = 4,5-3,4$ transition at $\nu$ = 279511.8577 MHz. This estimate takes into account the convolution of the map with the same beam width as in \cite{2011A&A...533A..85L} (35$^{\prime\prime}$). For the N$_2$H$^+$ (1--0) line, the overlap is not so high, only three central components overlap. Modeling shows that for the first clump, the overlap of three components with a line width of 4.55 km/s \citep{2011A&A...533A..85L} and a maximum intensity of the central component ($F_1F = 2,3-1,2$) of 1.1 K gives an observed intensity of 2.2 K. For the second clump with a line width is 3.06 km/s a maximum intensity of the most intense component  is 0.7 K, for the third clump it is 0.6 K. This analysis shows that the linewidth of N$_2$H$^+$ $J = 3-2$ line is higher than $J = 1-0$. It is likely that in the $J = 3-2$ line we see a denser gas with a higher turbulence. 

\cite{2011A&A...533A..85L} determined the N$_2$H$^+$ column density for the first clump 5.5--10.3 $\times 10^{13}$ cm$^{-2}$, for the second clump 2.5 $\times 10^{13}$ cm$^{-2}$ and for the fifth clump 1.8 $\times$ 10$^{13}$ cm$^{-2}$, in our analysis under the LTE assumption the column density of the first clump is 1.7 $\times$ 10$^{13}$ cm$^{-2}$, for the second clump it is 1.4 $\times$ 10$^{13}$ cm$^{-2}$, and for the third clump it is 1.1 $\times$ 10$^{13}$ cm$^{-2}$. A non-LTE analysis using the RADEX software of the N$_2$H$^+$ $J = 1-0$ and $J=3-2$ data shows that for the first clump the hydrogen density $n({\mathrm{H_2}}) \sim 3\times 10^{5}$ cm$^{-3}$ at the gas kinetic temperature $T_{\rm kin} \sim 30-100$~K with the column density $N(\mathrm{N_2H^+}) \approx 2-3 \times 10^{13}$~cm$^{-2}$, the model and observational data achieve the best agreement at $T_{\rm kin} \sim 40-50$~K and $N(\mathrm{N_2H^+}) \approx 2.5 \times 10^{13}$~cm$^{-2}$. For the second and the third  clumps  $n({\mathrm{H_2}}) \sim  5\times 10^{5}$ cm$^{-3}$ at $T_{\rm kin} \sim 15-30$~K with the averaged value of the column density of $N(\mathrm{N_2H^+})  \approx 2 \times 10^{13}$~cm$^{-2}$ for the second clump and  $N(\mathrm{N_2H^+}) \approx  10^{13}$~cm$^{-2}$ for the third clump.

\begin{table}
	\caption{  {Derived parameters of the hyperfine structure of the N$_2$H$^+$ (3--2) line for the different clumps}}
	\label{n2h_hpf_par}
	%\bigskip
	\begin{tabular}{ccc}
		\hline
		Clump & $T (F_1F = 4,5-3,4)$ & $\Delta V$ \\
		&(K) &(\kms) \\
		\hline
		1 & 0.401 $\pm$ 0.001 & 6.33 $\pm$ 0.02   \\
		2 & 0.361 $\pm$ 0.002 & 5.65 $\pm$ 0.03   \\
		3 & 0.219 $\pm$  0.002 & 5.16 $\pm$ 0.06 \\
		4 & 0.196 $\pm$ 0.003 & 4.12 $\pm$ 0.07 \\
		5 & 0.149 $\pm$ 0.001 & 4.92 $\pm$ 0.05   \\
		6 & 0.117 $\pm$ 0.002 & 3.47 $\pm$ 0.05 \\
		[1mm]
		\hline
	\end{tabular}
\end{table}

The HNCO data imply even higher densities in the first clump. As shown in \cite{Zinchenko2000} the $K_{-1}=0$ transitions can be excited by collisions at densities $n\ga 10^6$~\cmc. A collisional excitation of the $K_{-1}>0$ transitions requires very high densities. Most probably they are excited by Far-IR radiation, but densities in the emission regions should still be quite high. The emission in these transitions should arise in the close vicinity of the luminous IR source IRAS 17233--3606, which is consistent with the observed compactness of this emission. This clump was a target for many studies as mentioned above. We see there a range of temperatures from $\sim 25$~K derived from the \textit{Herschel} data to $\sim 300$~K from our HNCO data. Apparently, this reflects the fact that it contains a hot core surrounded by a much colder extended envelope. The shift of the high-excitation HNCO lines in velocity relative to the lower excitation lines cannot be explained in a spherically-symmetric or cylindrically-symmetric optically thin case. One possibility is to abandon the assumption of a low optical depth in the high-excitation HNCO lines. In this case we can try to attribute the velocity shift to a self-absorption in infalling outer layers of the hot core. However, the spectra presented in Fig.~\ref{fig:hnco-spectra} do not show signs of self-absorption and a high optical depth in so highly excited lines seems unrealistic and has never been observed in similar objects \citep{Zinchenko2000}. Another possibility is to assume an asymmetric distribution of the high-excitation HNCO molecules. In this case the shift can be attributed to orbital or radial motions. It is worth noting that \cite{Beuther2017} observed the blue-shifted absorption in the CH$_3$CN high-excitation transitions at about the same velocities as the high-excitation HNCO lines in our data. They attribute this shift to a contribution from the outflow. In this picture we have to assume that the high-excitation HNCO emission arises exclusively in the blue-shifted outflow lobe and is absent at the systemic velocity. Such assumption is rather strange and unusual. A more natural explanation is a movement of the hot dense core relative the surrounding medium at a velocity of a few \kms\ along the line of sight.
Such supersonic movements of young massive stars are frequently suggested to explain the morphology of UC \Hii\ regions \citep[e.g.][]{Wood1989}. Relative movements of a dense and more diffuse molecular gas have been reported, too, although at lower velocities \citep[e.g.][]{Kirsanova2008,Henshaw2013}. In principle such a movement hints at the scenario of a triggered star formation under the influence of an external factor, such as an expanding shell or cloud collision.

\begin{figure}
	\includegraphics[width=\linewidth]{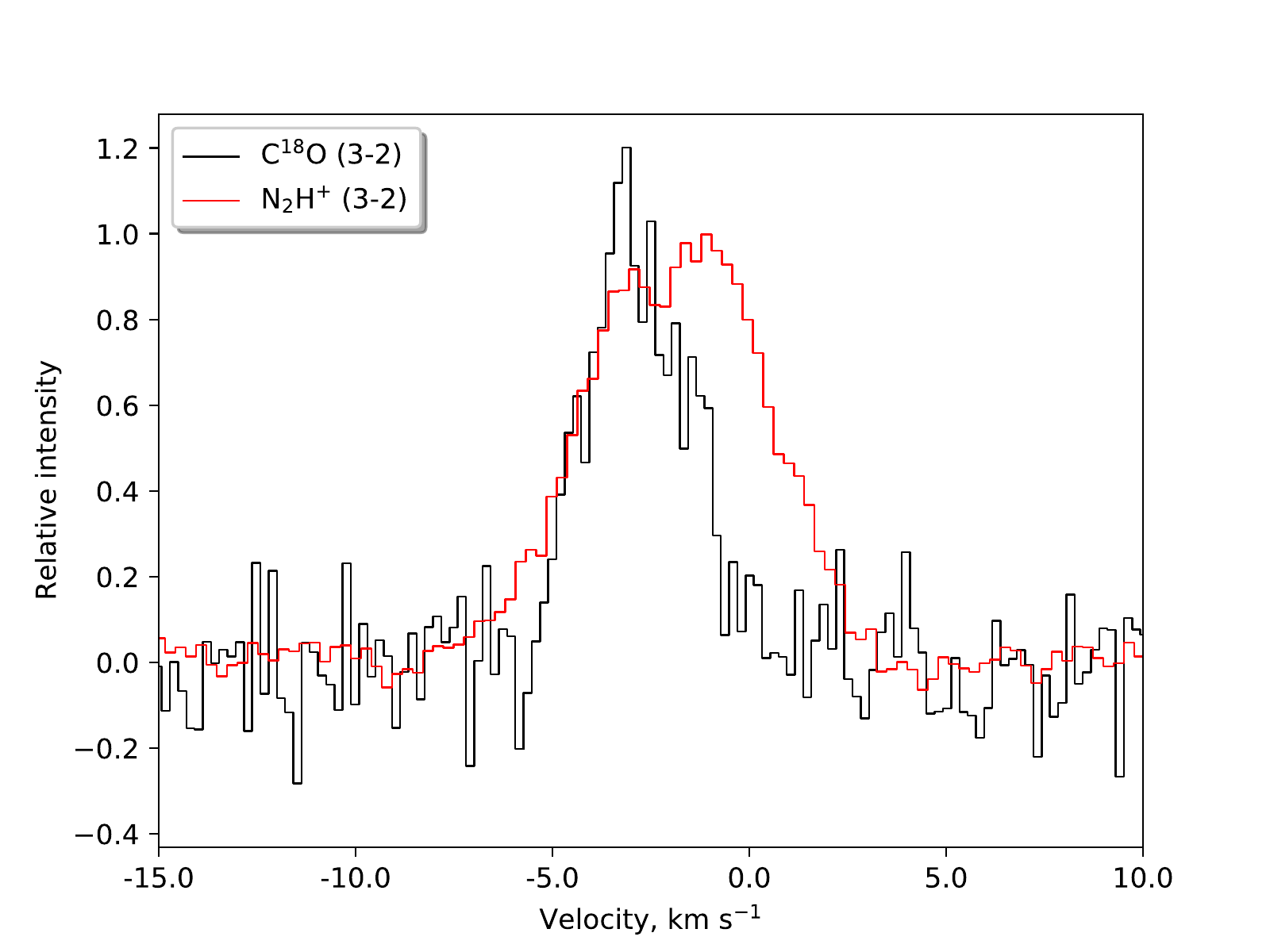} 
	\caption{An overlay of the C$^{18}$O (3-2) and N$_2$H$^+$ (3--2) spectra toward the third clump. The normalized intensities of the lines are used.}
	\label{fig:c18o+n2h}
\end{figure}

As noticed in Sect.~\ref{clump_sect} there is a velocity shift between N$_2$H$^+$ (3--2) and CO isotopologues in some clumps. Is is especially prominent in the third and forth clumps but is also noticeable in the sixth clump. Fig.~\ref{fig:c18o+n2h} shows this shift in the third clump very clearly. The difference in the velocities is $\sim 1$~\kms. A very similar picture is observed in the forth clump. This shift is consistent with such scenario, too. \cite{2011A&A...533A..85L} suggested that an external agent might be the cause of the line broadening along the filament. The velocity shifts observed in our data support this suggestion.

Widths of all observed lines are highly supersonic. The N$_2$H$^+$ (3--2) line width is 1.2--2.1 times higher than the C$^{18}$O line width in all clumps and than the N$_2$H$^+$ (1--0) line width in the clumps where it was observed. This hardly can be explained by the optical depth broadening \citep[e.g.][]{Phillips1979} since the brightness of the N$_2$H$^+$ (3--2) line and our modeling do not support an assumption of a high optical depth. Taking into account the fact that the critical density of the N$_2$H$^+$ (3--2) transition is much higher than critical density of the other transitions considered here, we can conclude  that the velocity dispersion increases in the central denser regions of the clumps. This increase hardly can be attributed to turbulence since most clumps lack powerful sources in their interiors. Most probably the line widths increase due to Keplerian-like rotation and/or infall motions in the clumps. As mentioned above the clumps are associated with mid-IR sources, which indicates their protostellar nature. The maps of the first moment in the N$_2$H$^+$ (3--2) line (Fig.~\ref{fig:moments}) show velocity gradients in the clumps which can indicate a rotation. In the first clump the gradient is similar to that observed at small scales \citep{Klaassen2015}, which was interpreted as an evidence for rotation. 
The sharp increase of the line width in the center of the clumps is consistent with Keplerian-like rotation, too. However, a rotation may be not a unique interpretation of the velocity gradients in some cases. In the northern part where we see a significant difference between the velocities of N$_2$H$^+$ (3--2) and CO isotopologues, the observed gradient can probably arise under the influence of an external factor. High resolution observations are needed to clarify kinematics of this area.

The N$_2$H$^+$ abundance is close to the typical values \citep[e.g.][]{Pirogov03}. Fig.~\ref{Nn2h} shows N$_2$H$^+$ depletion in the direction of the first clump, which contains the luminous IR source. Such behavior is rather typical for massive cores and can be probably explained by the dissociative recombination of N$_2$H$^+$ \citep{Pirogov07, Zin09}.

\section{Conclusions}
We performed a multi-line study of the filamentary infrared dark cloud G351.78--0.54 with the APEX radio telescope. The observed lines include CO (2--1), $^{13}$CO (2--1), C$^{18}$O (2--1), $^{13}$CO (3--2), 	C$^{18}$O (3--2),  N$_2$H$^+$ (3--2), CH$_3$CCH ($13_K-12_K$) and several HNCO transitions. The main results are the following:

\noindent
1. The main filamentary body was mapped in the CO (2--1), $^{13}$CO (2--1), C$^{18}$O (2--1), $^{13}$CO (3--2), 	C$^{18}$O (3--2) and  N$_2$H$^+$ (3--2) lines. Maps of the C$^{18}$O and N$_2$H$^+$ column densities have been constructed in the LTE approximation. The total mass of the filament is estimated at $\sim 1800$~M\sun. The mass per unit length ($M_{\rm line}$ = 529 M$_\odot$/pc) is close to the critical value. However, both values represent upper limits. The presence of several dense clumps along the filament shows that the process of fragmentation is going on.

\noindent
2. Six dense clumps are identified in the N$_2$H$^+$ (3--2) map. Their masses and virial parameters have been derived from the C$^{18}$O (2--1) data assuming gas temperature equal dust temperature obtained from the \textit{Herschel} data. All clumps except one appear gravitationally unstable. For two clumps we obtained temperature estimates from the CH$_3$CCH rotation diagrams. These temperatures are somewhat higher than the dust temperatures. In the first clump which contains the luminous IR source IRAS 17233-3606, the CH$_3$CCH rotation temperature is about 120~K. We use the C$^{18}$O (3--2)/(2--1) intensity ratio for estimation of the clump density on the basis of non-LTE modeling. The densities obtained in this way are $n \sim 3\times 10^4$~\cmc.  For the three clumps with the available N$_2$H$^+$ (1--0) data \citep{2011A&A...533A..85L} we estimate the density from the N$_2$H$^+$ (3--2)/(1--0) intensity ratio (taking into account the hyperfine splitting of these lines). These density estimates are about an order of magnitude higher, $n \sim 3\times 10^5$~\cmc, which is consistent with the fact that N$_2$H$^+$ traces denser regions than C$^{18}$O. The width of the N$_2$H$^+$ (3--2) lines is larger than the width of the N$_2$H$^+$ (1--0) and C$^{18}$O lines, which indicates a higher velocity dispersion in the denser parts of the clumps, which is most probably related to a Keplerian-like rotation or infall motions.

\noindent
3. Exclusively toward the first clump we detected several HNCO lines with the excitation energy of the upper level up to $\sim 800$~K. The emission in the $K_{-1}=0$ transitions is rather extended and the spectra show broad wings indicative of the outflow. The orientation of the outflow lobes is in a good agreement with the observations of the outflowing gas in the lines of other molecules. The HNCO data imply even higher density in this clump, $n\ga 10^6$~\cmc. The rotational temperature derived from the higher excitation transitions is $\sim 300$~K. There is a shift in velocity between the higher and lower excitation transitions, which most likely hints at a movement of the hot dense core relative the surrounding medium at a velocity of a few \kms\ along the line of sight.

\noindent
4. In some clumps there is a velocity shift $\sim 1$~\kms\ between N$_2$H$^+$ (3--2) and CO isotopologues. It indicates a relative movement of the dense and more diffuse gas and can be caused by an external agent as suggested earlier for explanation of the general line broadening in the filament \citep{2011A&A...533A..85L}.

\noindent
5. The N$_2$H$^+$ abundance is close to the typical values in general but drops toward the luminous IR source IRAS 17233--3606 in the first clump. This behavior is consistent with other similar objects.

\section{DATA AVAILABILITY}

Data directly related to this publication are available by request from the corresponding author. 

\section*{Acknowledgements}
We are very grateful to the anonymous referee for the constructive comments.
This research was supported by the Russian Foundation for Basic Research (grant No. 18-02-00660, initial data reduction) and by the Russian Science Foundation (grant No. 17-12-01256, data analysis).
Based on observations with the Atacama Pathfinder EXperiment (APEX) telescope. APEX is a collaboration between the Max Planck Institute for Radio Astronomy, the European Southern Observatory, and the Onsala Space Observatory. Swedish observations on APEX are supported through Swedish Research Council grant No 2017-00648.
\addcontentsline{toc}{section}{Acknowledgements}



\bibliographystyle{mnras}
\bibliography{example} % if your bibtex file is called example.bib

\begin{thebibliography}{}
\makeatletter
\relax
\def\mn@urlcharsother{\let\do\@makeother \do\$\do\&\do\#\do\^\do\_\do\%\do\~}
\def\mn@doi{\begingroup\mn@urlcharsother \@ifnextchar [ {\mn@doi@}
  {\mn@doi@[]}}
\def\mn@doi@[#1]#2{\def\@tempa{#1}\ifx\@tempa\@empty \href
  {http://dx.doi.org/#2} {doi:#2}\else \href {http://dx.doi.org/#2} {#1}\fi
  \endgroup}
\def\mn@eprint#1#2{\mn@eprint@#1:#2::\@nil}
\def\mn@eprint@arXiv#1{\href {http://arxiv.org/abs/#1} {{\tt arXiv:#1}}}
\def\mn@eprint@dblp#1{\href {http://dblp.uni-trier.de/rec/bibtex/#1.xml}
  {dblp:#1}}
\def\mn@eprint@#1:#2:#3:#4\@nil{\def\@tempa {#1}\def\@tempb {#2}\def\@tempc
  {#3}\ifx \@tempc \@empty \let \@tempc \@tempb \let \@tempb \@tempa \fi \ifx
  \@tempb \@empty \def\@tempb {arXiv}\fi \@ifundefined
  {mn@eprint@\@tempb}{\@tempb:\@tempc}{\expandafter \expandafter \csname
  mn@eprint@\@tempb\endcsname \expandafter{\@tempc}}}

\bibitem[\protect\citeauthoryear{{Andr{\'e}} et~al.,}{{Andr{\'e}}
  et~al.}{2010}]{Andre10}
{Andr{\'e}} P.,  et~al., 2010, \mn@doi [\aap] {10.1051/0004-6361/201014666},
  \href {https://ui.adsabs.harvard.edu/abs/2010A&A...518L.102A} {518, L102}

\bibitem[\protect\citeauthoryear{{Andr{\'e}}, {Di Francesco}, {Ward-Thompson},
  {Inutsuka}, {Pudritz}  \& {Pineda}}{{Andr{\'e}} et~al.}{2014}]{Andre2014}
{Andr{\'e}} P.,  {Di Francesco} J.,  {Ward-Thompson} D.,  {Inutsuka} S.-I.,
  {Pudritz} R.~E.,   {Pineda} J.~E.,  2014, \mn@doi [Protostars and Planets VI]
  {10.2458/azu_uapress_9780816531240-ch002}, \href
  {http://adsabs.harvard.edu/abs/2014prpl.conf...27A} {pp 27--51}

\bibitem[\protect\citeauthoryear{{Antyufeyev}, {Shulga}  \&
  {Zinchenko}}{{Antyufeyev} et~al.}{2016}]{Antyufeyev2016}
{Antyufeyev} O.~V.,  {Shulga} V.~M.,   {Zinchenko} I.~I.,  2016, \mn@doi
  [Kinematics and Physics of Celestial Bodies] {10.3103/S0884591316060027},
  \href {https://ui.adsabs.harvard.edu/abs/2016KPCB...32..276A} {32, 276}

\bibitem[\protect\citeauthoryear{{Astropy Collaboration} et~al.,}{{Astropy
  Collaboration} et~al.}{2018}]{2018AJ....156..123A}
{Astropy Collaboration} et~al., 2018, \mn@doi [\aj] {10.3847/1538-3881/aabc4f},
  \href {https://ui.adsabs.harvard.edu/abs/2018AJ....156..123A} {156, 123}

\bibitem[\protect\citeauthoryear{{Ballesteros-Paredes}, {Hartmann},
  {V{\'a}zquez-Semadeni}, {Heitsch}  \&
  {Zamora-Avil{\'e}s}}{{Ballesteros-Paredes}
  et~al.}{2011}]{2011MNRAS.411...65B}
{Ballesteros-Paredes} J.,  {Hartmann} L.~W.,  {V{\'a}zquez-Semadeni} E.,
  {Heitsch} F.,   {Zamora-Avil{\'e}s} M.~A.,  2011, \mn@doi [\mnras]
  {10.1111/j.1365-2966.2010.17657.x}, \href
  {https://ui.adsabs.harvard.edu/abs/2011MNRAS.411...65B} {411, 65}

\bibitem[\protect\citeauthoryear{{Bally}, {Langer}, {Stark}  \&
  {Wilson}}{{Bally} et~al.}{1987}]{Bally87}
{Bally} J.,  {Langer} W.~D.,  {Stark} A.~A.,   {Wilson} R.~W.,  1987, \mn@doi
  [\apjl] {10.1086/184817}, \href
  {https://ui.adsabs.harvard.edu/abs/1987ApJ...312L..45B} {312, L45}

\bibitem[\protect\citeauthoryear{{Battersby} et~al.,}{{Battersby}
  et~al.}{2011}]{2011A&A...535A.128B}
{Battersby} C.,  et~al., 2011, \mn@doi [\aap] {10.1051/0004-6361/201116559},
  \href {https://ui.adsabs.harvard.edu/abs/2011A&A...535A.128B} {535, A128}

\bibitem[\protect\citeauthoryear{{Belitsky} et~al.,}{{Belitsky}
  et~al.}{2006}]{2006SPIE.6275E..0GB}
{Belitsky} V.,  et~al., 2006, in Society of Photo-Optical Instrumentation
  Engineers (SPIE) Conference Series. p. 62750G, \mn@doi{10.1117/12.671383}

\bibitem[\protect\citeauthoryear{{Bergin}, {Goldsmith}, {Snell}  \&
  {Ungerechts}}{{Bergin} et~al.}{1994}]{Bergin1994}
{Bergin} E.~A.,  {Goldsmith} P.~F.,  {Snell} R.~L.,   {Ungerechts} H.,  1994,
  \mn@doi [\apj] {10.1086/174518}, \href
  {http://adsabs.harvard.edu/abs/1994ApJ...431..674B} {431, 674}

\bibitem[\protect\citeauthoryear{{Beuther}, {Walsh}, {Johnston}, {Henning},
  {Kuiper}, {Longmore}  \& {Walmsley}}{{Beuther} et~al.}{2017}]{Beuther2017}
{Beuther} H.,  {Walsh} A.~J.,  {Johnston} K.~G.,  {Henning} T.,  {Kuiper} R.,
  {Longmore} S.~N.,   {Walmsley} C.~M.,  2017, \mn@doi [\aap]
  {10.1051/0004-6361/201630126}, \href
  {https://ui.adsabs.harvard.edu/abs/2017A&A...603A..10B} {603, A10}

\bibitem[\protect\citeauthoryear{{Beuther} et~al.,}{{Beuther}
  et~al.}{2019}]{Beuther2019}
{Beuther} H.,  et~al., 2019, \mn@doi [\aap] {10.1051/0004-6361/201834064},
  \href {https://ui.adsabs.harvard.edu/abs/2019A&A...621A.122B} {621, A122}

\bibitem[\protect\citeauthoryear{{Contreras} et~al.,}{{Contreras}
  et~al.}{2013}]{Contreras2013}
{Contreras} Y.,  et~al., 2013, \mn@doi [\aap] {10.1051/0004-6361/201220155},
  \href {https://ui.adsabs.harvard.edu/abs/2013A&A...549A..45C} {549, A45}

\bibitem[\protect\citeauthoryear{{Currie}, {Berry}, {Jenness}, {Gibb}, {Bell}
  \& {Draper}}{{Currie} et~al.}{2014}]{Currie2014}
{Currie} M.~J.,  {Berry} D.~S.,  {Jenness} T.,  {Gibb} A.~G.,  {Bell} G.~S.,
  {Draper} P.~W.,  2014, in {Manset} N.,  {Forshay} P.,  eds,  Astronomical
  Society of the Pacific Conference Series Vol. 485, Astronomical Data Analysis
  Software and Systems XXIII. p.~391

\bibitem[\protect\citeauthoryear{{Dewangan}, {Pirogov}, {Ryabukhina}, {Ojha}
  \& {Zinchenko}}{{Dewangan} et~al.}{2019}]{2019ApJ...877....1D}
{Dewangan} L.~K.,  {Pirogov} L.~E.,  {Ryabukhina} O.~L.,  {Ojha} D.~K.,
  {Zinchenko} I.,  2019, \mn@doi [\apj] {10.3847/1538-4357/ab1aa6}, \href
  {https://ui.adsabs.harvard.edu/abs/2019ApJ...877....1D} {877, 1}

\bibitem[\protect\citeauthoryear{{Endres}, {Schlemmer}, {Schilke}, {Stutzki}
  \& {M{\"u}ller}}{{Endres} et~al.}{2016}]{Endres16}
{Endres} C.~P.,  {Schlemmer} S.,  {Schilke} P.,  {Stutzki} J.,   {M{\"u}ller}
  H. S.~P.,  2016, \mn@doi [Journal of Molecular Spectroscopy]
  {10.1016/j.jms.2016.03.005}, \href
  {https://ui.adsabs.harvard.edu/abs/2016JMoSp.327...95E} {327, 95}

\bibitem[\protect\citeauthoryear{{Ginsburg} \& {Mirocha}}{{Ginsburg} \&
  {Mirocha}}{2011}]{2011ascl.soft09001G}
{Ginsburg} A.,  {Mirocha} J.,  2011, {PySpecKit: Python Spectroscopic Toolkit}
  (\mn@eprint {ascl} {1109.001})

\bibitem[\protect\citeauthoryear{{Goldsmith} \& {Langer}}{{Goldsmith} \&
  {Langer}}{1999}]{Goldsmith1999}
{Goldsmith} P.~F.,  {Langer} W.~D.,  1999, \mn@doi [\apj] {10.1086/307195},
  \href {https://ui.adsabs.harvard.edu/abs/1999ApJ...517..209G} {517, 209}

\bibitem[\protect\citeauthoryear{{G{\"u}sten}, {Nyman}, {Schilke}, {Menten},
  {Cesarsky}  \& {Booth}}{{G{\"u}sten} et~al.}{2006}]{2006A&A...454L..13G}
{G{\"u}sten} R.,  {Nyman} L.~{\AA}.,  {Schilke} P.,  {Menten} K.,  {Cesarsky}
  C.,   {Booth} R.,  2006, \mn@doi [\aap] {10.1051/0004-6361:20065420}, \href
  {https://ui.adsabs.harvard.edu/abs/2006A%26A...454L..13G} {454, L13}

\bibitem[\protect\citeauthoryear{{Henshaw}, {Caselli}, {Fontani},
  {Jim{\'e}nez-Serra}, {Tan}  \& {Hernandez}}{{Henshaw}
  et~al.}{2013}]{Henshaw2013}
{Henshaw} J.~D.,  {Caselli} P.,  {Fontani} F.,  {Jim{\'e}nez-Serra} I.,  {Tan}
  J.~C.,   {Hernandez} A.~K.,  2013, \mn@doi [\mnras] {10.1093/mnras/sts282},
  \href {https://ui.adsabs.harvard.edu/abs/2013MNRAS.428.3425H} {428, 3425}

\bibitem[\protect\citeauthoryear{{Kauffmann} \& {Pillai}}{{Kauffmann} \&
  {Pillai}}{2010}]{Kauffmann2010}
{Kauffmann} J.,  {Pillai} T.,  2010, \mn@doi [\apjl]
  {10.1088/2041-8205/723/1/L7}, \href
  {https://ui.adsabs.harvard.edu/abs/2010ApJ...723L...7K} {723, L7}

\bibitem[\protect\citeauthoryear{{Kauffmann}, {Pillai}  \&
  {Goldsmith}}{{Kauffmann} et~al.}{2013}]{Kauffmann2013}
{Kauffmann} J.,  {Pillai} T.,   {Goldsmith} P.~F.,  2013, \mn@doi [\apj]
  {10.1088/0004-637X/779/2/185}, \href
  {http://adsabs.harvard.edu/abs/2013ApJ...779..185K} {779, 185}

\bibitem[\protect\citeauthoryear{{Kirsanova}, {Sobolev}, {Thomasson}, {Wiebe},
  {Johansson}  \& {Seleznev}}{{Kirsanova} et~al.}{2008}]{Kirsanova2008}
{Kirsanova} M.~S.,  {Sobolev} A.~M.,  {Thomasson} M.,  {Wiebe} D.~S.,
  {Johansson} L.~E.~B.,   {Seleznev} A.~F.,  2008, \mn@doi [\mnras]
  {10.1111/j.1365-2966.2008.13429.x}, \href
  {https://ui.adsabs.harvard.edu/abs/2008MNRAS.388..729K} {388, 729}

\bibitem[\protect\citeauthoryear{{Klaassen}, {Johnston}, {Leurini}  \&
  {Zapata}}{{Klaassen} et~al.}{2015}]{Klaassen2015}
{Klaassen} P.~D.,  {Johnston} K.~G.,  {Leurini} S.,   {Zapata} L.~A.,  2015,
  \mn@doi [\aap] {10.1051/0004-6361/201424781}, \href
  {https://ui.adsabs.harvard.edu/abs/2015A&A...575A..54K} {575, A54}

\bibitem[\protect\citeauthoryear{{Koumpia}, {Harvey}, {Ossenkopf}, {van der
  Tak}, {Mookerjea}, {Fuente}  \& {Kramer}}{{Koumpia}
  et~al.}{2015}]{Koumpia2015}
{Koumpia} E.,  {Harvey} P.~M.,  {Ossenkopf} V.,  {van der Tak} F.~F.~S.,
  {Mookerjea} B.,  {Fuente} A.,   {Kramer} C.,  2015, \mn@doi [\aap]
  {10.1051/0004-6361/201525669}, \href
  {https://ui.adsabs.harvard.edu/abs/2015A&A...580A..68K} {580, A68}

\bibitem[\protect\citeauthoryear{{Leurini}, {Hieret}, {Thorwirth}, {Wyrowski},
  {Schilke}, {Menten}, {G{\"u}sten}  \& {Zapata}}{{Leurini}
  et~al.}{2008}]{Leurini2008}
{Leurini} S.,  {Hieret} C.,  {Thorwirth} S.,  {Wyrowski} F.,  {Schilke} P.,
  {Menten} K.~M.,  {G{\"u}sten} R.,   {Zapata} L.,  2008, \mn@doi [\aap]
  {10.1051/0004-6361:200809475}, \href
  {https://ui.adsabs.harvard.edu/abs/2008A&A...485..167L} {485, 167}

\bibitem[\protect\citeauthoryear{{Leurini}, {Codella}, {Zapata}, {Beltr{\'a}n},
  {Schilke}  \& {Cesaroni}}{{Leurini} et~al.}{2011a}]{Leurini2011}
{Leurini} S.,  {Codella} C.,  {Zapata} L.,  {Beltr{\'a}n} M.~T.,  {Schilke} P.,
    {Cesaroni} R.,  2011a, \mn@doi [\aap] {10.1051/0004-6361/201016190}, \href
  {https://ui.adsabs.harvard.edu/abs/2011A&A...530A..12L} {530, A12}

\bibitem[\protect\citeauthoryear{{Leurini}, {Pillai}, {Stanke}, {Wyrowski},
  {Testi}, {Schuller}, {Menten}  \& {Thorwirth}}{{Leurini}
  et~al.}{2011b}]{2011A&A...533A..85L}
{Leurini} S.,  {Pillai} T.,  {Stanke} T.,  {Wyrowski} F.,  {Testi} L.,
  {Schuller} F.,  {Menten} K.~M.,   {Thorwirth} S.,  2011b, \mn@doi [\aap]
  {10.1051/0004-6361/201016380}, \href
  {http://adsabs.harvard.edu/abs/2011A%26A...533A..85L} {533, A85}

\bibitem[\protect\citeauthoryear{{Leurini} et~al.,}{{Leurini}
  et~al.}{2014}]{Leurini2014}
{Leurini} S.,  et~al., 2014, \mn@doi [\aap] {10.1051/0004-6361/201323343},
  \href {https://ui.adsabs.harvard.edu/abs/2014A&A...564L..11L} {564, L11}

\bibitem[\protect\citeauthoryear{{Leurini} et~al.,}{{Leurini}
  et~al.}{2019}]{2019A&A...621A.130L}
{Leurini} S.,  et~al., 2019, \mn@doi [\aap] {10.1051/0004-6361/201833612},
  \href {http://adsabs.harvard.edu/abs/2019A%26A...621A.130L} {621, A130}

\bibitem[\protect\citeauthoryear{{Li}, {Urquhart}, {Leurini}, {Csengeri},
  {Wyrowski}, {Menten}  \& {Schuller}}{{Li} et~al.}{2016}]{Li16}
{Li} G.-X.,  {Urquhart} J.~S.,  {Leurini} S.,  {Csengeri} T.,  {Wyrowski} F.,
  {Menten} K.~M.,   {Schuller} F.,  2016, \mn@doi [\aap]
  {10.1051/0004-6361/201527468}, \href
  {http://cdsads.u-strasbg.fr/abs/2016A%26A...591A...5L} {591, A5}

\bibitem[\protect\citeauthoryear{{Liu}, {Wu}  \& {Zhang}}{{Liu}
  et~al.}{2013}]{Liu2013}
{Liu} T.,  {Wu} Y.,   {Zhang} H.,  2013, \mn@doi [\apjl]
  {10.1088/2041-8205/775/1/L2}, \href
  {https://ui.adsabs.harvard.edu/abs/2013ApJ...775L...2L} {775, L2}

\bibitem[\protect\citeauthoryear{{Low} et~al.,}{{Low} et~al.}{1984}]{Low84}
{Low} F.~J.,  et~al., 1984, \mn@doi [\apjl] {10.1086/184213}, \href
  {https://ui.adsabs.harvard.edu/abs/1984ApJ...278L..19L} {278, L19}

\bibitem[\protect\citeauthoryear{{Mallick}, {Ojha}, {Tamura}, {Linz}, {Samal}
  \& {Ghosh}}{{Mallick} et~al.}{2015}]{2015MNRAS.447.2307M}
{Mallick} K.~K.,  {Ojha} D.~K.,  {Tamura} M.,  {Linz} H.,  {Samal} M.~R.,
  {Ghosh} S.~K.,  2015, \mn@doi [\mnras] {10.1093/mnras/stu2584}, \href
  {https://ui.adsabs.harvard.edu/abs/2015MNRAS.447.2307M} {447, 2307}

\bibitem[\protect\citeauthoryear{{Mangum} \& {Shirley}}{{Mangum} \&
  {Shirley}}{2015}]{Mangum2015}
{Mangum} J.~G.,  {Shirley} Y.~L.,  2015, \mn@doi [\pasp] {10.1086/680323},
  \href {http://adsabs.harvard.edu/abs/2015PASP..127..266M} {127, 266}

\bibitem[\protect\citeauthoryear{{Maret}, {Hily-Blant}, {Pety}, {Bardeau}  \&
  {Reynier}}{{Maret} et~al.}{2011}]{Maret2011}
{Maret} S.,  {Hily-Blant} P.,  {Pety} J.,  {Bardeau} S.,   {Reynier} E.,  2011,
  \mn@doi [\aap] {10.1051/0004-6361/201015487}, \href
  {http://adsabs.harvard.edu/abs/2011A%26A...526A..47M} {526, A47}

\bibitem[\protect\citeauthoryear{{McClure-Griffiths}, {Dickey}, {Gaensler},
  {Green}  \& {Haverkorn}}{{McClure-Griffiths} et~al.}{2006}]{McClure06}
{McClure-Griffiths} N.~M.,  {Dickey} J.~M.,  {Gaensler} B.~M.,  {Green} A.~J.,
   {Haverkorn} M.,  2006, \mn@doi [\apj] {10.1086/508706}, \href
  {https://ui.adsabs.harvard.edu/abs/2006ApJ...652.1339M} {652, 1339}

\bibitem[\protect\citeauthoryear{{Motte}, {Bontemps}  \& {Louvet}}{{Motte}
  et~al.}{2018}]{Motte18}
{Motte} F.,  {Bontemps} S.,   {Louvet} F.,  2018, \mn@doi [\araa]
  {10.1146/annurev-astro-091916-055235}, \href
  {https://ui.adsabs.harvard.edu/abs/2018ARA&A..56...41M} {56, 41}

\bibitem[\protect\citeauthoryear{{M{\"u}ller}, {Thorwirth}, {Roth}  \&
  {Winnewisser}}{{M{\"u}ller} et~al.}{2001}]{Mueller01}
{M{\"u}ller} H.~S.~P.,  {Thorwirth} S.,  {Roth} D.~A.,   {Winnewisser} G.,
  2001, \mn@doi [\aap] {10.1051/0004-6361:20010367}, \href
  {https://ui.adsabs.harvard.edu/abs/2001A&A...370L..49M} {370, L49}

\bibitem[\protect\citeauthoryear{{M{\"u}ller}, {Schl{\"o}der}, {Stutzki}  \&
  {Winnewisser}}{{M{\"u}ller} et~al.}{2005}]{Mueller05}
{M{\"u}ller} H.~S.~P.,  {Schl{\"o}der} F.,  {Stutzki} J.,   {Winnewisser} G.,
  2005, \mn@doi [Journal of Molecular Structure]
  {10.1016/j.molstruc.2005.01.027}, \href
  {http://cdsads.u-strasbg.fr/abs/2005JMoSt.742..215M} {742, 215}

\bibitem[\protect\citeauthoryear{{Myers}}{{Myers}}{2009}]{Myers09}
{Myers} P.~C.,  2009, \mn@doi [\apj] {10.1088/0004-637X/700/2/1609}, \href
  {https://ui.adsabs.harvard.edu/#abs/2009ApJ...700.1609M} {700, 1609}

\bibitem[\protect\citeauthoryear{{Pagani}, {Daniel}  \& {Dubernet}}{{Pagani}
  et~al.}{2009}]{2009A&A...494..719P}
{Pagani} L.,  {Daniel} F.,   {Dubernet} M.~L.,  2009, \mn@doi [\aap]
  {10.1051/0004-6361:200810570}, \href
  {https://ui.adsabs.harvard.edu/abs/2009A&A...494..719P} {494, 719}

\bibitem[\protect\citeauthoryear{{Phillips}, {Huggins}, {Wannier}  \&
  {Scoville}}{{Phillips} et~al.}{1979}]{Phillips1979}
{Phillips} T.~G.,  {Huggins} P.~J.,  {Wannier} P.~G.,   {Scoville} N.~Z.,
  1979, \mn@doi [\apj] {10.1086/157237}, \href
  {https://ui.adsabs.harvard.edu/abs/1979ApJ...231..720P} {231, 720}

\bibitem[\protect\citeauthoryear{{Pickett}, {Poynter}, {Cohen}, {Delitsky},
  {Pearson}  \& {M{\"u}ller}}{{Pickett} et~al.}{1998}]{Pickett98}
{Pickett} H.~M.,  {Poynter} R.~L.,  {Cohen} E.~A.,  {Delitsky} M.~L.,
  {Pearson} J.~C.,   {M{\"u}ller} H.~S.~P.,  1998, \mn@doi [\jqsrt]
  {10.1016/S0022-4073(98)00091-0}, \href
  {http://cdsads.u-strasbg.fr/abs/1998JQSRT..60..883P} {60, 883}

\bibitem[\protect\citeauthoryear{{Pilbratt} et~al.,}{{Pilbratt}
  et~al.}{2010}]{Pilbratt10}
{Pilbratt} G.~L.,  et~al., 2010, \mn@doi [\aap] {10.1051/0004-6361/201014759},
  \href {http://cdsads.u-strasbg.fr/abs/2010A%26A...518L...1P} {518, L1}

\bibitem[\protect\citeauthoryear{{Pirogov}, {Zinchenko}, {Caselli}, {Johansson}
   \& {Myers}}{{Pirogov} et~al.}{2003}]{Pirogov03}
{Pirogov} L.,  {Zinchenko} I.,  {Caselli} P.,  {Johansson} L.~E.~B.,   {Myers}
  P.~C.,  2003, \mn@doi [\aap] {10.1051/0004-6361:20030659}, \href
  {http://cdsads.u-strasbg.fr/abs/2003A%26A...405..639P} {405, 639}

\bibitem[\protect\citeauthoryear{{Pirogov}, {Zinchenko}, {Caselli}  \&
  {Johansson}}{{Pirogov} et~al.}{2007}]{Pirogov07}
{Pirogov} L.,  {Zinchenko} I.,  {Caselli} P.,   {Johansson} L.~E.~B.,  2007,
  \mn@doi [\aap] {10.1051/0004-6361:20054777}, \href
  {http://adsabs.harvard.edu/abs/2007A%26A...461..523P} {461, 523}

\bibitem[\protect\citeauthoryear{{Rathborne}, {Jackson}  \&
  {Simon}}{{Rathborne} et~al.}{2006}]{Rathborne06}
{Rathborne} J.~M.,  {Jackson} J.~M.,   {Simon} R.,  2006, \mn@doi [\apj]
  {10.1086/500423}, \href {http://cdsads.u-strasbg.fr/abs/2006ApJ...641..389R}
  {641, 389}

\bibitem[\protect\citeauthoryear{{Rathborne} et~al.,}{{Rathborne}
  et~al.}{2016}]{2016PASA...33...30R}
{Rathborne} J.~M.,  et~al., 2016, \mn@doi [\pasa] {10.1017/pasa.2016.23}, \href
  {https://ui.adsabs.harvard.edu/abs/2016PASA...33...30R} {33, e030}

\bibitem[\protect\citeauthoryear{{Ryabukhina}, {Zinchenko}, {Samal},
  {Zemlyanukha}, {Ladeyschikov}, {Sobolev}, {Henkel}  \& {Ojha}}{{Ryabukhina}
  et~al.}{2018}]{Ryabukhina2018}
{Ryabukhina} O.~L.,  {Zinchenko} I.~I.,  {Samal} M.~R.,  {Zemlyanukha} P.~M.,
  {Ladeyschikov} D.~A.,  {Sobolev} A.~M.,  {Henkel} C.,   {Ojha} D.~K.,  2018,
  \mn@doi [Research in Astronomy and Astrophysics] {10.1088/1674-4527/18/8/95},
  \href {https://ui.adsabs.harvard.edu/abs/2018RAA....18...95R} {18, 095}

\bibitem[\protect\citeauthoryear{{Sault}, {Teuben}  \& {Wright}}{{Sault}
  et~al.}{1995}]{Sault1995}
{Sault} R.~J.,  {Teuben} P.~J.,   {Wright} M.~C.~H.,  1995, in {Shaw} R.~A.,
  {Payne} H.~E.,   {Hayes} J.~J.~E.,  eds,  Astronomical Society of the Pacific
  Conference Series Vol. 77, Astronomical Data Analysis Software and Systems
  IV. p.~433 (\mn@eprint {} {astro-ph/0612759})

\bibitem[\protect\citeauthoryear{{Schisano} et~al.,}{{Schisano}
  et~al.}{2020}]{Schisano20}
{Schisano} E.,  et~al., 2020, \mn@doi [\mnras] {10.1093/mnras/stz3466}, \href
  {https://ui.adsabs.harvard.edu/abs/2020MNRAS.492.5420S} {492, 5420}

\bibitem[\protect\citeauthoryear{{Snell} \& {Loren}}{{Snell} \&
  {Loren}}{1977}]{Snell1977}
{Snell} R.~L.,  {Loren} R.~B.,  1977, \mn@doi [\apj] {10.1086/154909}, \href
  {https://ui.adsabs.harvard.edu/abs/1977ApJ...211..122S} {211, 122}

\bibitem[\protect\citeauthoryear{{Stutzki} \& {Guesten}}{{Stutzki} \&
  {Guesten}}{1990}]{Stutzki1990}
{Stutzki} J.,  {Guesten} R.,  1990, \mn@doi [\apj] {10.1086/168859}, \href
  {http://adsabs.harvard.edu/abs/1990ApJ...356..513S} {356, 513}

\bibitem[\protect\citeauthoryear{{Tan}, {Beltr{\'a}n}, {Caselli}, {Fontani},
  {Fuente}, {Krumholz}, {McKee}  \& {Stolte}}{{Tan} et~al.}{2014}]{Tan14}
{Tan} J.~C.,  {Beltr{\'a}n} M.~T.,  {Caselli} P.,  {Fontani} F.,  {Fuente} A.,
  {Krumholz} M.~R.,  {McKee} C.~F.,   {Stolte} A.,  2014, \mn@doi [Protostars
  and Planets VI] {10.2458/azu_uapress_9780816531240-ch007}, \href
  {http://adsabs.harvard.edu/abs/2014prpl.conf..149T} {pp 149--172}

\bibitem[\protect\citeauthoryear{{Van der Tak}, {Black}, {Sch{\"o}ier},
  {Jansen}  \& {van Dishoeck}}{{Van der Tak}
  et~al.}{2007}]{2007A&A...468..627V}
{Van der Tak} F.~F.~S.,  {Black} J.~H.,  {Sch{\"o}ier} F.~L.,  {Jansen} D.~J.,
   {van Dishoeck} E.~F.,  2007, \mn@doi [\aap] {10.1051/0004-6361:20066820},
  \href {https://ui.adsabs.harvard.edu/abs/2007A%26A...468..627V} {468, 627}

\bibitem[\protect\citeauthoryear{{Vassilev} et~al.,}{{Vassilev}
  et~al.}{2008}]{2008A&A...490.1157V}
{Vassilev} V.,  et~al., 2008, \mn@doi [\aap] {10.1051/0004-6361:200810459},
  \href {https://ui.adsabs.harvard.edu/abs/2008A%26A...490.1157V} {490, 1157}

\bibitem[\protect\citeauthoryear{{Wienen} et~al.,}{{Wienen}
  et~al.}{2015}]{2015A&A...579A..91W}
{Wienen} M.,  et~al., 2015, \mn@doi [\aap] {10.1051/0004-6361/201424802}, \href
  {http://adsabs.harvard.edu/abs/2015A%26A...579A..91W} {579, A91}

\bibitem[\protect\citeauthoryear{{Wood} \& {Churchwell}}{{Wood} \&
  {Churchwell}}{1989}]{Wood1989}
{Wood} D. O.~S.,  {Churchwell} E.,  1989, \mn@doi [\apjs] {10.1086/191329},
  \href {https://ui.adsabs.harvard.edu/abs/1989ApJS...69..831W} {69, 831}

\bibitem[\protect\citeauthoryear{{Yu}, {Xu}, {Wang}  \& {Liu}}{{Yu}
  et~al.}{2018}]{Yu2018}
{Yu} N.-P.,  {Xu} J.-L.,  {Wang} J.-J.,   {Liu} X.-L.,  2018, \mn@doi [\apj]
  {10.3847/1538-4357/aadb94}, \href
  {https://ui.adsabs.harvard.edu/abs/2018ApJ...865..135Y} {865, 135}

\bibitem[\protect\citeauthoryear{{Zinchenko}, {Henkel}  \& {Mao}}{{Zinchenko}
  et~al.}{2000}]{Zinchenko2000}
{Zinchenko} I.,  {Henkel} C.,   {Mao} R.~Q.,  2000, \aap, \href
  {https://ui.adsabs.harvard.edu/abs/2000A&A...361.1079Z} {361, 1079}

\bibitem[\protect\citeauthoryear{{Zinchenko}, {Caselli}  \&
  {Pirogov}}{{Zinchenko} et~al.}{2009}]{Zin09}
{Zinchenko} I.,  {Caselli} P.,   {Pirogov} L.,  2009, \mn@doi [\mnras]
  {10.1111/j.1365-2966.2009.14687.x}, \href
  {http://adsabs.harvard.edu/abs/2009MNRAS.395.2234Z} {395, 2234}

\makeatother
\end{thebibliography}

%\begin{thebibliography}{99}
%\end{thebibliography}

%%%%%%%%%%%%%%%%%%%%%%%%%%%%%%%%%%%%%%%%%%%%%%%%%%

%%%%%%%%%%%%%%%%% APPENDICES %%%%%%%%%%%%%%%%%%%%%

% Don't change these lines
\bsp	% typesetting comment
\label{lastpage}
\end{document}